\documentclass[a4paper,fleqn]{cas-sc}

\usepackage[numbers,sort&compress]{natbib}

\usepackage{tikz}
\usepackage{amsmath,amssymb}
\usepackage{colortbl}
\usepackage{xcolor}
\usepackage{tcolorbox}
\usepackage{listings}
\usepackage{url}
\usepackage{xspace}
\usepackage{ifthen}
\usepackage[T1]{fontenc}
\usepackage[inline]{enumitem}
\usepackage[framemethod=tikz]{mdframed}
\usepackage{booktabs}   %% for \toprule / \midrule / \bottomrule in tables
\usepackage{caption}
\usepackage[section]{placeins}

\hypersetup{breaklinks=true}

\lstdefinestyle{privacycode}{
  language=,
  basicstyle=\ttfamily\footnotesize,
  keywordstyle=\color{blue!80!black}\bfseries,
  commentstyle=\color{green!60!black}\itshape,
  stringstyle=\color{red!80!black},
  numberstyle=\tiny\color{gray},
  numbers=left,
  stepnumber=1,
  numbersep=5pt,
  backgroundcolor=\color{gray!5},
  frame=single,
  rulecolor=\color{gray!30},
  breaklines=true
}

\newboolean{showcomments}
\setboolean{showcomments}{false}

\ifthenelse{\boolean{showcomments}}
  {\newcommand{\nb}[2]{
   \fbox{\bfseries\sffamily\scriptsize#1}
   {\sf\small$\blacktriangleright$\textit{\textcolor{red}{#2}}$\blacktriangleleft$}
  }}
  {\newcommand{\nb}[2]{}
   
  }

\newcommand{\toolName}{PolicyGapper\xspace}
\newcommand{\toolLink}{\url{https://github.com/Mobile-IoT-Security-Lab/PolicyGapper}\xspace}

\begin{document}
\let\WriteBookmarks\relax
\def\floatpagepagefraction{1}
\def\textpagefraction{.001}

%% Running header — short title and authors
\shorttitle{PolicyGapper: Automated Detection of PP--DSS Inconsistencies}
\shortauthors{Ferrari et al.}

%% Full title
\title[mode=title]{PolicyGapper: Automated Detection of Inconsistencies
Between Google Play Data Safety Sections and Privacy Policies Using LLMs}

%% -------------------------------------------------------
%% Authors and affiliations
%% -------------------------------------------------------

%% --- Author 1 ---
\author[1]{Luca Ferrari}
\cormark[1]
\ead{luca.ferrari@imtlucca.it}
\credit{Conceptualization, Methodology, Software, Validation, Writing -- original draft}

\affiliation[1]{
  organization={IMT School for Advanced Studies Lucca},
  addressline={Piazza San Francesco 19},
  city={Lucca},
  postcode={55100},
  country={Italy}
}
%% --- Author 4 ---
\author[2]{Billel Habbati}
\ead{9032167@studenti.unige.it}
\credit{Conceptualization, Writing -- review \& editing}

%% --- Author 5 ---
\author[2]{Meriem Guerar}
\ead{meriem.guerar@unige.it}
\credit{Conceptualization, Writing -- review \& editing}

%% --- Author 3 ---
\author[3]{Mariano Ceccato}
\ead{mariano.ceccato@univr.it}
\credit{Conceptualization, Methodology, Writing -- review \& editing, Supervision}

\affiliation[3]{
  organization={University of Verona},
  addressline={Strada le Grazie 15},
  city={Verona},
  postcode={37134},
  country={Italy}
}
%% --- Author 2 ---
\author[2]{Luca Verderame}
\ead{luca.verderame@unige.it}
\credit{Conceptualization, Methodology, Writing -- review \& editing, Supervision}

\affiliation[2]{
  organization={DIBRIS, University of Genova},
  addressline={Via Dodecaneso 35},
  city={Genova},
  postcode={16146},
  country={Italy}
}

%% Corresponding author footnote
\cortext[1]{Corresponding author}

%% -------------------------------------------------------
%% Abstract
%% -------------------------------------------------------
\begin{abstract}
Mobile application developers are required to disclose how they collect, use, and share user data in compliance with privacy regulations. To support transparency, major app marketplaces have introduced standardized disclosure mechanisms. In 2022, Google mandated the Data Safety Section (DSS) on Google Play, requiring developers to summarize their data practices. However, compiling accurate DSS disclosures is challenging, as they must remain consistent with the corresponding privacy policy (PP), and no automated tool currently verifies this alignment. Prior studies indicate that nearly 80\% of popular apps contain incomplete or misleading DSS declarations.

We present \toolName, an LLM-based methodology for automatically detecting discrepancies between DSS disclosures and privacy policies. \toolName operates in four stages: scraping, pre-processing, analysis, and post-processing, without requiring access to application binaries. We evaluate \toolName on a dataset of 330 top-ranked apps spanning all 33 Google Play categories, collected in Q3 2025. The approach identifies 2,689 omitted disclosures, including 2,040 related to data collection and 649 to data sharing. Manual validation on a stratified 10\% subset, repeated across three independent runs, yields an average Precision of 0.75, Recall of 0.77, Accuracy of 0.69, and F1-score of 0.76. To support reproducibility, we release a complete replication package, including the dataset, prompts, source code, and results available at \url{https://github.com/Mobile-IoT-Security-Lab/PolicyGapper} and \url{https://doi.org/10.5281/zenodo.19628493}.
\end{abstract}

%% -------------------------------------------------------
%% Highlights (required by Elsevier; 3--5 items, ≤85 chars each)
%% -------------------------------------------------------
\begin{highlights}
\item First LLM-based pipeline to automatically audit PP--DSS consistency at scale
\item Detected 2,689 omitted declarations across 330 Google Play apps (Q3 2025)
\item Average F1-score of 0.76 on manually validated ground-truth subset
\end{highlights}

%% -------------------------------------------------------
%% Keywords
%% -------------------------------------------------------

\maketitle

%% -------------------------------------------------------
%% Sections — paths unchanged from USENIX version
%% Put all .tex files in a subfolder named sections/
%% -------------------------------------------------------

\section{Introduction}
\label{sec:intro}
The rapid growth of mobile applications has raised increasing concerns about how user data is collected, processed, and shared. Privacy policies remain the primary mechanism to ensure transparency, yet their effectiveness is often questioned. Several studies have shown that the Privacy Policy (PP) of mobile apps is complex, vague, contradictory, and often ineffective~\cite{limits}~\cite{short}~\cite{contraddiction}, which results in users not reading them. As a consequence, users remain unaware of how their personal data is handled, limiting their ability to provide informed consent and undermining the role of privacy policies as tools for transparency and accountability.
To facilitate user comprehension of privacy policies, Privacy labels introduced by Kelly et al.~\cite{labels} have been implemented in the tech industry.
These labels provide a standardized, graphical summary of an application's data practices---such as data collection, sharing, and usage purposes---allowing users to quickly understand key privacy information without parsing lengthy legal texts.

Following this paradigm, Google introduced the Data Safety Sections to help users comprehend complex app privacy practices~\cite{zhang2022usable}. As reported in the official documentation of the Google Play Store~\cite{GooglePlayDoc}, developers are asked to compile the DSS for each app describing all the collection, sharing, and other practices for a range of 14 different user data categories, as well as the purposes for which they use those data in accordance with the corresponding PP.
In this context, data collection refers to any instance in which the application gathers user data (e.g., during account creation or app usage), while data sharing indicates the transfer of such data to third parties.

Unfortunately, recent studies~\cite{mozilla,datarequests} have revealed that, in most cases, mobile apps fail to report data types or operations that should be declared according to the DSS requirements. Even more concerning, a 2025 report from NowSecure \cite{nowsecure2025}  found that 10\% of the analyzed Android apps do not publish a Data Safety Section on the Google Play Store, in clear violation of the store’s regulations.

These omissions or incomplete disclosures can have a twofold impact. On the one hand, users may develop a false sense of privacy regarding how their personal data is handled, undermining their ability to make informed decisions about app usage and contradicting the transparency goals of the DSS. On the other hand, developers may incur fines for non-compliance with applicable data privacy and data protection laws (e.g., up to 4\% of the annual worldwide turnover~\cite{penality}), while Google may reject updates or even remove non-compliant apps from the store~\cite{googlePenality}.

Despite the importance and central role of the DSS in ensuring transparency and regulatory compliance, verifying its consistency with the information stated in the PP remains a challenging task. Manual assessment of these documents requires both technical and legal expertise. It involves analyzing actual data behavior and interpreting privacy statements to evaluate compliance with store policies and applicable regulations. In addition, such a process is time-consuming, resource-intensive, and often beyond the capabilities of individual developers or small organizations.
Moreover, the evolution of applications through the release of updates could introduce misalignment between the different versions of the DSS and the PP, making the identification of these inconsistencies even more difficult.

In this context, automated tools that support developers in verifying the consistency between the DSS and the corresponding PP would be highly beneficial. However, to the best of our knowledge, neither the existing literature nor the Google Play Store provides systematic or automated solutions for this task.

To bridge this gap, we propose a novel assessment methodology that exploits a multi-prompt Large Language Models (LLMs) pipeline to i) extract all the declarations, defined as textual statements that explicitly describe data collection or sharing practices associated with specific user data, from both the privacy policy and the DSS, and ii) identify the \emph{omitted declarations}, i.e., declarations contained in the PP but not covered in the corresponding DSS.
In addition, we implemented the methodology in a fully-automated tool, called \textbf{\toolName}, to detect such omissions in Android apps published on the Google Play Store.

Leveraging \toolName, we conducted a study to evaluate the consistency of the Android privacy ecosystem and assess the effectiveness of our approach. Specifically, we addressed the following research questions:

\begin{itemize}
    \item \textbf{RQ1:} Can \toolName effectively identify omitted declarations between the PP and the DSS?
    \item \textbf{RQ2:} Which types of omitted declarations appear most frequently during the analysis?
    \item \textbf{RQ3:} Which Data Categories are most impacted by omitted declarations among the Google Play Store app Categories?
\end{itemize}

To answer these questions, we tested \toolName on 330 top apps, collected in Q3 2025, across all the available categories of the Google Play Store. \toolName was able to report \textbf{2,689 issues}, corresponding to omitted \textit{Data Types} declared in the PP but not in the DSS, revealing that more than 95\% of the apps analyzed contained verified omitted declarations.
%We manually reviewed 10\% of the PPs to validate the results and assess Precision, Accuracy, Recall, and F1-score, obtaining the following results in an average of 3 runs: Precision of 0.75, Accuracy of 0.69, Recall of 0.77, and F1-score of 0.76. 
We manually reviewed 10\% of the privacy policies to validate the results. Across three runs, the average Precision is 0.75, Accuracy 0.69, Recall 0.77, and F1-score 0.76.

%At the end of the experimental assessment, in the second half of January, we notified the developers of these apps via email to share our findings;  this allowed developers to update their DSS, making it consistent with what is written in the PP.
%Despite the short time that has passed, seven developers from different companies have already responded, accepting the report that we sent and highlighting their attention to these issues.
%We plan to disclose our methodology's performance and results to Google, potentially informing future enhancements to their app review process.

At the end of the experimental assessment, in the second half of January, we responsibly disclosed our findings to the developers of the affected applications via email. The goal of this disclosure was to inform developers of the identified inconsistencies and enable corrective actions, allowing them to update the DSS to better reflect the practices described in the PP.
Within the limited time elapsed since disclosure, some developers from different companies have already responded, acknowledging our report.

We further plan to share aggregated results and methodological insights with Google, with the objective of informing potential improvements to the Google Play app review and compliance processes. Lastly, to promote replicability and adoption of the tool and methodology, we released the open-source tool, the used datasets, and results at the following link\footnote{\toolLink , \url{https://doi.org/10.5281/zenodo.19628493}}.

The rest of the paper is organized as follows. Section~\ref{sec:bg} provides the necessary background, followed by a review of related work in Section~\ref{sec:rw}. Section~\ref{sec:methodology} details our proposed methodology, while Section~\ref{sec:implementation} describes the system implementation. Section~\ref{sec:ExpRes} presents the experimental evaluation and analyzes the results. Section~\ref{Discussion}  discusses the results obtained and the limitations of \toolName. Finally, Section~\ref{sec:Conclusion} summarizes our contributions and outlines directions for future research.

\section{Background}
\label{sec:bg}

This section summarizes the terminology and concepts required to understand the content of the rest of the paper.

Throughout this paper we adopt Google's terminology for data types, data categories, and disclosure constraints, as defined in the official Google Play Data Safety documentation.

%\textcolor{red}{Please note that we use Google's terminology}

\subsection{Google Data Safety Section.}
\label{GDSS}
The DSS is the standardized framework used by the Google Play Store to disclose how applications handle user data. It appears as a dedicated ``Data safety'' panel on each app's detail page. Developers complete this panel before submitting the app for review. They specify the data categories collected or shared (e.g., personal information, location, app activity), along with their data types and purposes (e.g., app functionality, analytics, advertising). They must also declare whether security practices are implemented, such as encryption in transit and user-requested data deletion. This information is displayed to users before installation (as shown in Figure~\ref{fig:dss}).
% \lv{qua sopra manca a ribadire che è lo sviluppatore che la deve fare, che sono privacy labels (vedi un pò i concetti espressi nell'intro e nell'abstract}
The DSS structures privacy-related information across a four-layer hierarchy:  Privacy Practices, Data Categories, Data Types, and Purposes~\cite{hierarchyDss}.
Developers must specify each layer when declaring data collection or sharing. They select the applicable privacy practice, identify the data categories involved, specify the data types collected, and declare the processing purposes. % \lv{qua però bisogna dare intuizione che il developer deve specidicare la practice, il dato che è interassato (data categories), etc..}

At the first level, as shown in Figure~\ref{fig:dssH}, Google Play Store defines three main Privacy Practices: 

\textbf{i) Data Collection.} According to Google’s official documentation~\cite{GooglePlayDoc}, ``collection'' refers to any transmission of user data from an app to a destination outside the user’s device. Data transmitted by integrated SDKs or libraries must always be reported as collected, regardless of whether it is sent to the developer’s or a third party’s server. 
When the app controls the code executed within a WebView (i.e., a browsing window integrated into the app), any user data gathered there must also be declared. In contrast, data produced from independent user browsing in windows outside the perimeter of the app (e.g., when a user opens a link in the browser of the mobile phone) does not require disclosure. 
% \lv{non è uncontrolled webview - anche perchè chi le controlla, cosa sono? - direi che è più l'attività di browsing in finestre al di fuori del perimetro dell'app, es quando l'utente apre un link nel browser del telefono invece non devono essere considerate}
% Temporary, in-memory data transfers used solely for real-time processing (i.e., data that is not stored or repurposed) are exempt. For example, a weather app that transmits the user’s location to fetch current conditions qualifies as ephemeral use. However, any data reused for profiling, advertising, or analytics is non-ephemeral and must be disclosed. Finally, even pseudonymous data must be reported as collected when it can reasonably be linked to a specific user.
Google Play explicitly excludes certain data access scenarios from the definition of ``collection'' when the data cannot be accessed by the developer or linked to a specific user. These exemptions include:
\begin{enumerate*}[label=(\alph*)]
    \item \textit{on-device processing}, where data is processed locally without transmission;
    \item \textit{end-to-end encryption}, where data is unreadable by the developer or intermediaries;
    \item \textit{ephemeral processing}, where data is briefly stored in memory for a real-time action and immediately discarded without being logged; and
    \item \textit{anonymization}, where data is stripped of all identifiers such that it can no longer be associated with an individual. In contrast, \textit{pseudonymization} retains a linkable ID and must still be declared.
\end{enumerate*} 

\textbf{ii) Data Sharing.}
Google Play’s Data Safety framework distinguishes between \textit{first} and \textit{third} parties. The \textit{first party} is the main entity responsible for processing data collected through the app—typically the organization listed as the developer on Google Play—and must ensure users clearly understand who handles their data. A \textit{third party} refers to any organization other than the first party or its designated service providers.
The term \textit{``sharing''} refers to the transfer of user data collected from an app to a third party, whether off-device or on-device. Off-device sharing occurs, for instance, when data is first transmitted to the developer’s server and subsequently forwarded to a third party. Sharing can also happen on-device, when data is transmitted directly from one app to another.
%—such cases must be disclosed in the app’s Data Safety Section (DSS), even if the data does not leave the device. 
Likewise, when third-party SDKs integrated into an app send user data directly to external servers, or when a WebView under the app’s control transmits user data, such activities constitute data sharing. 
By contrast, data exchanges resulting from independent user browsing in windows outside the perimeter of the app do not need to be disclosed. Certain data transfers fall outside the definition of sharing. These include transfers to a \textit{service provider} acting strictly on behalf of the developer and under its instructions; transfers conducted for \textit{legal obligations} or \textit{official requests}; user-initiated transfers or those made with clear in-app consent consistent with the \textit{User Data Policy}; and those involving fully anonymized data that cannot be linked to any identifiable individual. 

\textbf{iii) Security Practices.} Enable developers to showcase privacy protections through three optional tags: \textit{Encrypted in Transit}, \textit{Data Deletion Option}, and \textit{Review against Global Security Standards}. 

The second and third levels define the user data covered by the DSS, specifying what types of data are collected or shared by an application. In the current version, this information is organized into 14 high-level data categories, each further divided into 38 specific data types, providing a fine-grained taxonomy of user data as defined by Google Play~\cite{GooglePlayDoc}.
 % \lv{queste parti sono sbilanciate. qua dovresti dire che il 2 e il 3 livello mappano e definiscono i dati utente che sono oggetto della DSS; nella versione attuale ci sono 14 categorie, suddivise in 38 sotto categorie specifiche. aggiungere reference alla tabella}
The final level defines the Purposes, which describe the intended use of the collected or shared data and provide context on why the data is processed. These purposes include \textit{App functionality}, \textit{Analytics}, \textit{Developer communications}, \textit{Advertising or marketing}, \textit{Fraud prevention, security and compliance}, \textit{Personalization}, and \textit{Account management}.

\subsection{Google App Submission Review.}
\label{GASR}

In order to publish an app on the Google Play Store, developers are required to complete the DSS in the Google Play Console and provide a link to their PP.
As stated in Google Play documentation~\cite{GooglePlayDoc}, “even if your app does not collect any user data, you must still submit a form and provide a link to a privacy policy.”
Moreover, the PP URL must point to a stable, publicly accessible, read-only webpage (i.e., not publicly editable)~\cite{GooglePlayUserData}.
Finally, Google requires that DSS disclosures, compiled by developers, are consistent with the app’s privacy policy, i.e., the DSS should be aligned with the PP~\cite{GooglePlayDoc}.

To support developers in complying with these requirements, Google provides a set of tools integrated into the Play Console ecosystem. In particular, the Policy Status module~\cite{Policystatus} acts as a centralized dashboard for policy compliance monitoring, allowing developers to review enforcement actions (e.g., app rejections or removals) and track deadlines for newly introduced policy requirements. However, it primarily provides a reactive view of compliance issues after they have been detected, rather than proactively assisting developers in producing consistent documentation.

Despite its usefulness, this solutions do not fully address the DSS–PP consistency problem. In particular, while they support compliance monitoring and enforcement, they do not directly assess the semantic consistency between the structured DSS labels and the unstructured legal text of the PP. Consequently, an automated approach capable of interpreting natural language to audit and enforce alignment between PP and DSS remains necessary.

\subsection{Privacy Policy.}
In the mobile ecosystem, applications act as digital services subject to stringent data protection regulations (e.g., GDPR, CCPA) that mandate the transparent disclosure of data handling practices. Consequently, a PP is not merely a legal formality but a mandatory requirement for app distribution on platforms like the Google Play Store. This document legally binds the developer to specific standards, detailing the extent of data collection, the purposes of data processing, and third-party sharing, and serves as the authoritative reference for user consent and regulatory compliance.

\vspace{1em}
\noindent\begin{minipage}{\textwidth}
    \centering
    \begin{minipage}[c]{0.30\textwidth}
        \centering
        \includegraphics[width=\linewidth, keepaspectratio]{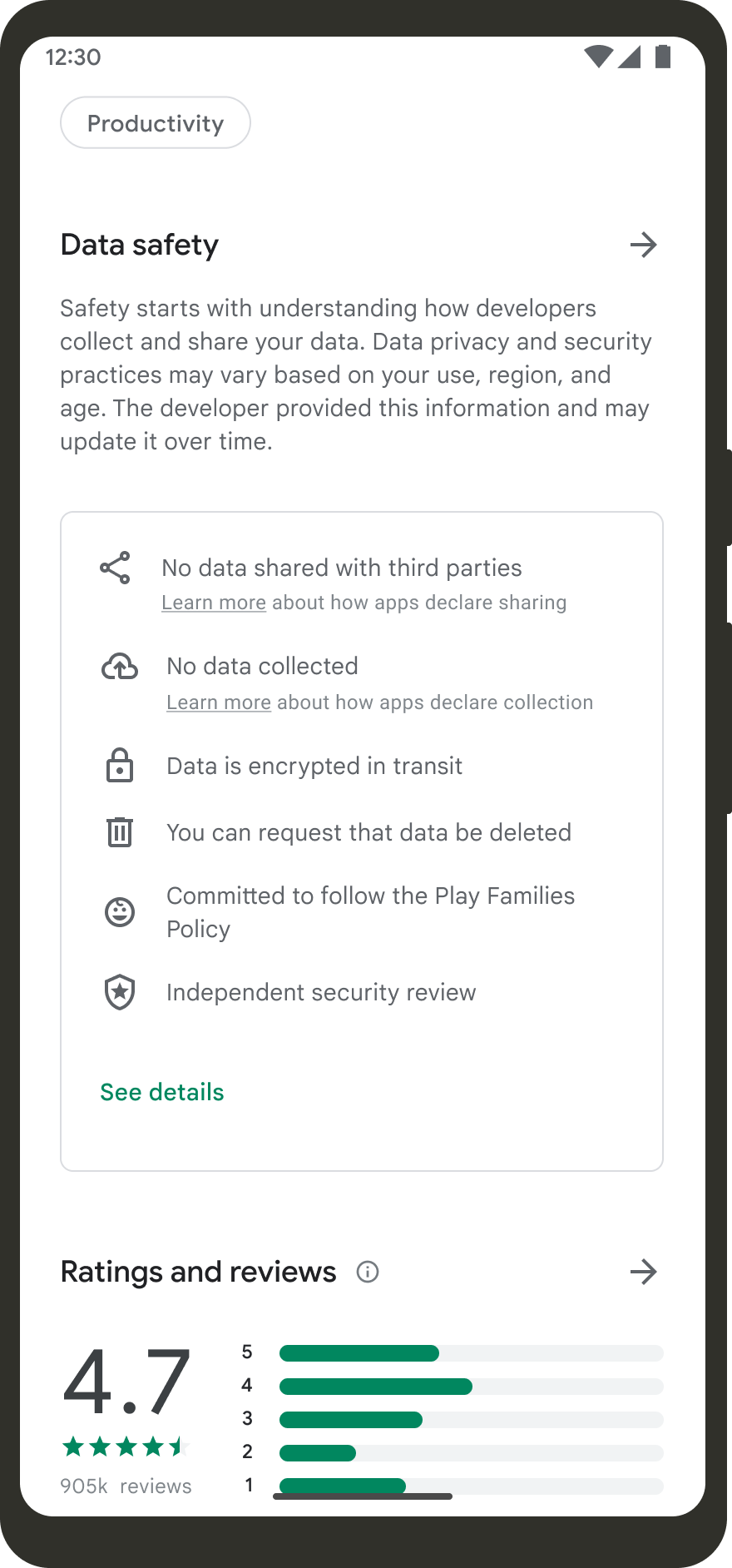}
        \captionof{figure}{Example of a Data Safety Section (DSS) for an application on the Google Play Store.}
        \label{fig:dss}
    \end{minipage}
    \hfill
    \begin{minipage}[c]{0.65\textwidth}
        \centering
        \includegraphics[width=\linewidth, keepaspectratio]{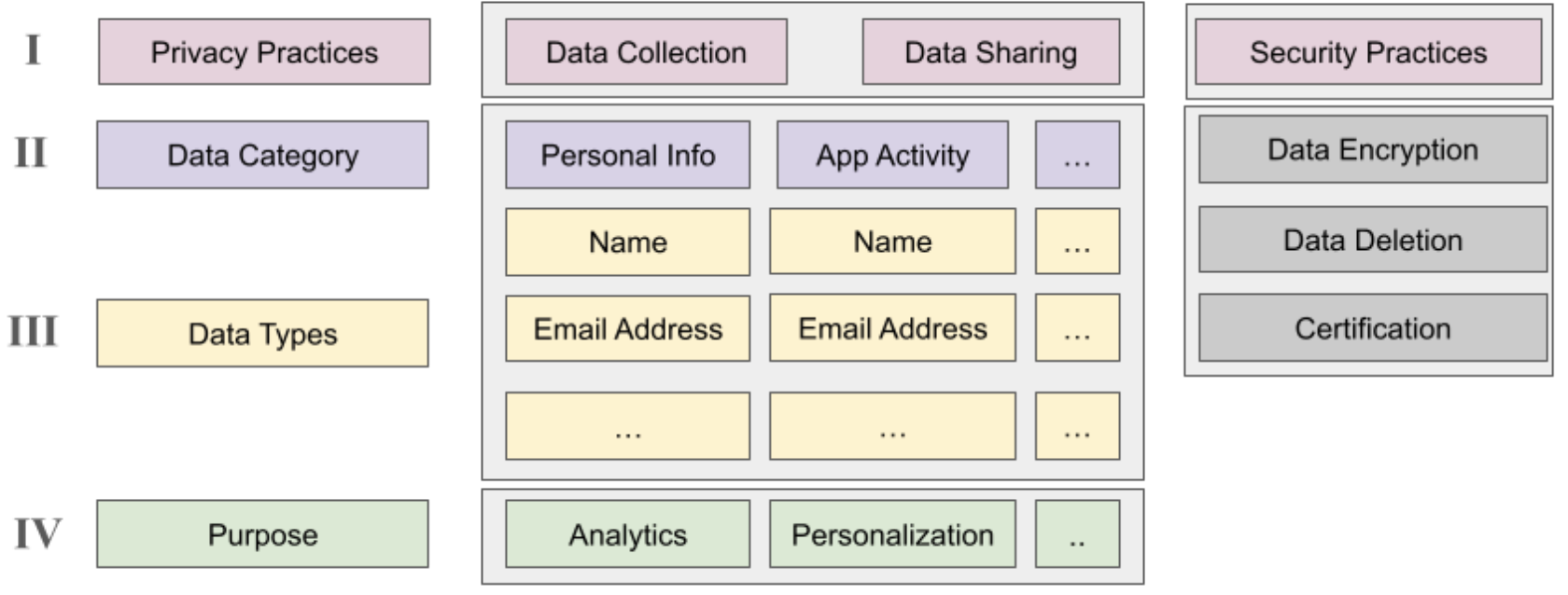}
        \captionof{figure}{Hierarchy of the Google Play Store Data Safety Section (DSS), adapted from Khandelwal et al.~\cite{hierarchyDss}.}
        \label{fig:dssH}
    \end{minipage}
\end{minipage}

\section{Related Work}
\label{sec:rw}

We categorize the related work into three primary domains: (i) approaches verifying consistency between app behavior and privacy documentation, (ii) single-artifact analyses of PP, and (iii) baseline measurements of privacy label correctness.

Regarding the first domain, several works focus on identifying discrepancies between code or app behaviour and PP, DSS, or GDPR requirements. In particular, Google provides Checks \cite{Checks}, a platform designed to support developers in assessing privacy compliance. However, the service is currently released in beta and has not received significant updates since May 2024, suggesting a limited level of maturity and maintenance. Checks operates as a separate service that can be connected to an application to perform automated analyses. The analysis is limited to applications published on the Google Play Store and accessible only to their respective developers. It inspects app characteristics such as permissions and network traffic to identify potential data collection and sharing behaviours, and applies natural language processing (NLP) techniques to analyze the Privacy Policy. Based on these analyses, it provides suggestions to assist developers in completing the Data Safety Section (DSS). However, its results are not part of the official Google Play review process and are not shared with the Play Store. Moreover, the tool does not explicitly perform a comparison between DSS disclosures and the PP, thus leaving the consistency verification task largely to developers.
3PDroid ~\cite{verderame2020reliability} verifies whether an application invokes APIs related to personal data collection or sharing before the user has given consent by accepting the PP.
Alecci et al.~\cite{alecci2025toward} employ LLMs to verify whether Android applications comply with privacy requirements derived from the GDPR by analyzing their Smali code.
Baalous et al.~\cite{baalous2025detecting} verify if the application's activity names are compliant with the privacy labels of the DSS.
GNChecker~\cite{fan2024giving} employ a combination of Static Application Security Testing (SAST), Dynamic Application Security Testing (DAST), and the use of LLMs.
While it innovatively uses LLMs to extract privacy statements, its primary goal is to validate these statements against technical data flows captured via static and dynamic analysis.
Similarly, Arkalakis et al.~\cite{arkalakis2024abandon} build an automated analysis framework that dynamically executes and analyzes applications so as to uncover discrepancies between the applications’ behavior and the data practices that
have been reported in their DSS.

Although these works are essential for detecting privacy leaks and contribute to improving transparency by identifying mismatches between application behavior and declared data practices, they address a different dimension of transparency violations. In particular, they focus on behavioral transparency, i.e., whether the actual data flows of an application are consistent with what is declared in the PP or DSS. 

However, they do not explicitly detect cross-document inconsistencies between the full description of data practices in the PP and the summarized disclosures in the DSS. As a result, even when an application’s behavior is correctly aligned with its declared practices, discrepancies between PP and DSS may still persist, potentially leading to incomplete or misleading disclosures for end users.

Moreover, these approaches face intrinsic scalability and accuracy challenges. Static analysis techniques can be hindered by anti-debugging and anti-repackaging mechanisms (e.g., code obfuscation and control-flow transformations), while dynamic analysis often suffers from limited code coverage. In practice, dynamic techniques (e.g., fuzzing) can only explore the portions of code that are actually executed, typically covering no more than 30\% of the application~\cite{akinotcho2025mobile}, which may result in missed sensitive data flows and false negatives.

The second domain focuses on single-artifact analysis, effectively identifying internal contradictions or legal gaps but failing to verify the cross-document alignment required by app stores.
For instance, Polisis~\cite{polisis} introduced a deep learning framework to extract fine-grained data practices from privacy policy text, allowing for automated querying and the generation of simplified visual summaries for users.
PolicyLint~\cite{contraddiction} analyzes over 11k apps to detect internal contradictions within Privacy Policies that suggest misleading statements. Similarly, Xie et al.~\cite{xie2025evaluating} use LLMs to evaluate website privacy policies against GDPR and US state laws. 
While these approaches ensure that %a single document (
the PP is internally consistent or legally compliant, they cannot detect transparency violations where the %legal document (
PP is misrepresented by the summarized disclosure in the DSS. Our system bridges this gap by performing a cross-verification between these two distinct artifacts.

Finally, the third domain involves baseline measurements of privacy labels. Khandelwal et al.~\cite{khandelwal2023overviewprivacylabelscompatibility} pioneered this area by examining over a million apps using BERT-based classifiers to detect potential inconsistencies. Subsequent works extended this line of work. One study tracked the longitudinal evolution of DSS, revealing widespread underreporting of ad/analytics SDKs~\cite{hierarchyDss}, while another identified significant discrepancies in self-reported practices across Android and iOS platforms~\cite{Khandelwal2}. However, these approaches rely on standard supervised learning, which limits their scope to predefined categories and keyword matching. This method often struggles with the ambiguous legal language found in modern policies. Our research advances this domain by replacing classical classification with the context-aware semantic reasoning of LLMs. This shift enables a deeper validation, identifying intricate discrepancies, such as conditional data sharing clauses, that escape conventional statistical methods. Furthermore, unlike prior studies, which did not release their artifacts, we provide open access to our tool and dataset to support reproducibility and future research.

\section{Methodology}
\label{sec:methodology}
We propose a novel methodology based on the combination of multi-prompt LLM analyses. The methodology is applicable to Android applications published on the Google Play Store without access to the application package or any additional information. It is composed of five cooperating modules, orchestrated according to the workflow depicted in Figure~\ref{fig:methodology}.

\vspace{1em}
\noindent\begin{minipage}{\textwidth}
    \centering
    \includegraphics[width=\textwidth]{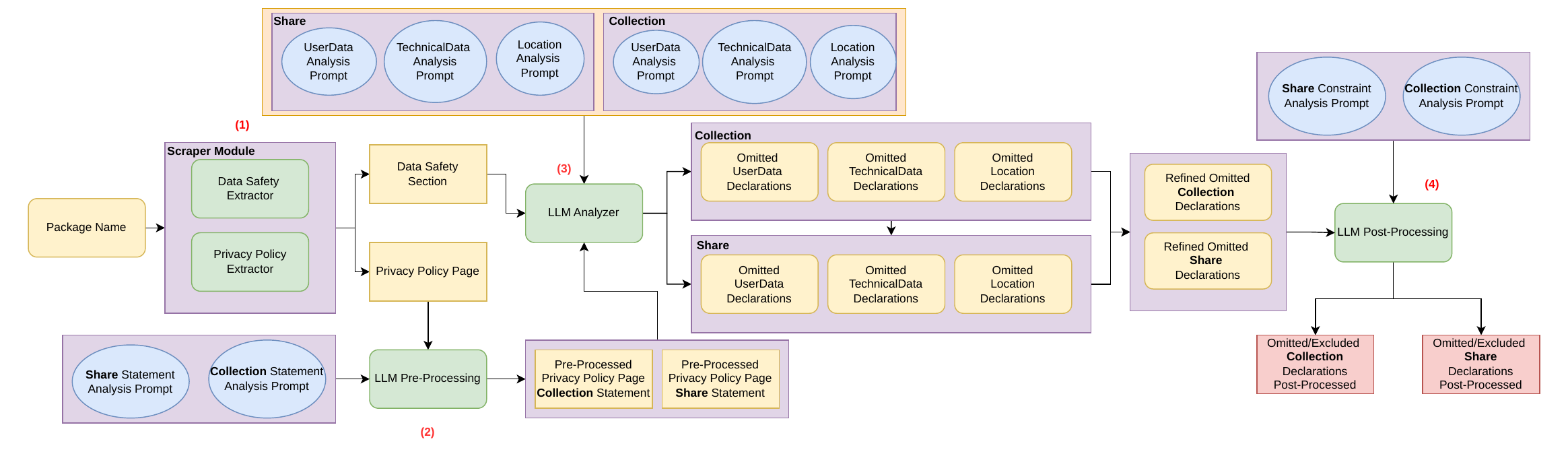}
    \captionof{figure}{Workflow of the proposed methodology for detecting inconsistencies between an app’s PP and DSS. }
    \label{fig:methodology}

    \end{minipage}
\paragraph{1) Scraper Module.}
The Scraper Module is responsible for retrieving the PP and the DSS associated with a given package name from the Google Play Store. It consists of two submodules: (i) the \textit{Data Safety Extractor}, which retrieves the DSS content, and (ii) the \textit{Privacy Policy Extractor}, which obtains the PP starting from the link provided on the DSS. 

% \lv{la metodologia la lascerei generica rispetto a PDF o non PDF altrimenti rischiamo di cadere nel punto che avevamo già deciso prima di decidere}

\paragraph{2) LLM Pre-Processing Module.}
%\textcolor{red}{manca assesmente della validazione del preprocessing}
Once the \textit{Privacy Policy Extractor} correctly extracts the PP of the given package name, 
the LLM Pre-Processing Module extracts the text statements of the PP that are relevant to data \textit{collection} and data \textit{sharing}. This module is executed through two dedicated prompts:
(i) the \textit{Share Statement Analysis Prompt} and
(ii) the \textit{Collection Statement Analysis Prompt}.
These two prompts are responsible for instructing the LLM to extract from the PP only the text blocks that describe the collection and sharing of personal user data (i.e., the declarations), and to copy them word-for-word without modification.
We introduce this module for two main reasons: i) PPs often span dozens of pages; isolating only the statements relevant to data collection and sharing practices of the mobile app significantly reduces the input size for the subsequent analysis, enabling a more focused and cost-effective LLM evaluation;
ii) the text extraction may yield poorly formatted outputs due to layout or encoding artifacts (e.g., broken symbols or misplaced characters). Through a dedicated preprocessing prompt, we transform these extracted sections into clean, interpretable text. As a result, the main analysis operates on compact and coherent inputs, minimizing noise and improving the overall robustness of the pipeline.
% \lv{forse ha senso mettere nella metdologia già i prompt; poi nell'implementazione raccontiamo che LLM abbiamo usato? altrimenti qua rimane un pò zoppo; questo vale anche per i sistemi sotto? conviene parlarne un attimo}

\paragraph{3) LLM Analyzer Module.}
%When the \textit{LLM Pre-Processing} module finishes its analysis, the \textit{LLM Analyzer} starts.
The LLM Analyzer is the core module responsible for comparing:
(i) the pre-processed PP statements related to data \textit{collection} and \textit{sharing}, and
(ii) the DSS extracted from Google Play. Its goal is to detect omitted declarations in the DSS, identifying cases in which specific data types are mentioned in the PP but not declared in the DSS.
The analysis relies on a set of structured prompts explicitly designed to guide the LLM in performing a compliance-oriented evaluation. Each prompt instructs the model to act as a privacy auditor specialized in Android applications, taking as input both the results of LLM Pre-Processing Module and the DSS. Depending on the data to analyze and for what purpose, i.e., \textit{collection} or \textit{sharing}, the LLM is prompted to identify all explicit or implicit mentions of specific \textit{data types} within the PP and to verify their corresponding declarations in the DSS. The module applies exclusion constraints derived from Google Play’s official documentation, as described in Section~\ref{GDSS}, to reduce false positives due to contextual or technical exceptions, such as on-device processing exemptions. The module considers all data types defined in the official Google Play Data Safety documentation. At the time of writing, the documentation specifies 38 \textit{Data Types} grouped into 14 \textit{Data Categories}.  

As output, the module produces two reports, summarizing potential omissions and inconsistencies separately for the \textit{collection} and \textit{sharing} practices. Each report covers all 14 data categories, providing a complete overview of undeclared or mismatched data practices. The output structure, as shown in Listing~\ref{lst:json-output} consists of the following elements:
\begin{itemize}
    \item \texttt{data\_type}: The candidate data type omitted in the DSS but present in the PP;
    \item \texttt{policy\_reference}: Exact excerpt from the privacy policy demonstrating the omission of the declared \texttt{data\_type} from the corresponding DSS entry;
    \item \texttt{lang}: language in which the PP is written.
\end{itemize}

\begin{lstlisting}[ caption={LLM Analyzer JSON output structure.}, label={lst:json-output},  style=privacycode, escapeinside={(*@}{@*)}, numbers=none]
{
    "omitted_declarations": [
        {
            "data_type": "Data type",
            "policy_reference": "Exact excerpt from the privacy policy",
            "lang": "PP language"
        }
    ]
}
\end{lstlisting}

Once the \textit{LLM Analyzer Module} completes its analysis, it consolidates the results into two unified artifacts: one for collection and one for sharing. This ensures that all candidate omissions are evaluated under a consistent set of constraints defined by the official Google Play Data Safety documentation, as described in Section~\ref{sec:bg}.

As output, the module returns:
(i) a single JSON file for potentially omitted \textit{sharing} declarations, and
(ii) a single JSON file for potentially omitted \textit{collection} declarations.
These files constitute the input for the post-processing phase.

\paragraph{5) LLM Post-Processing Module.}
The final stage of the pipeline is the \textit{LLM Post-Processing Module}.
This component refines and validates the outputs produced by the \textit{LLM Analyzer}, ensuring that each candidate omission represents a genuine inconsistency between the PP and the DSS.

We introduce this module to mitigate the known issues of hallucinations in LLMs' results~\cite{hallucination}, where models generate semantically plausible but factually inconsistent associations, such as linking unrelated data types like location sharing to payment info~\cite{farquhar2024}. The module operates through two dedicated prompts: (i) the \textit{Sharing Constraint Analysis Prompt}, and (ii) the \textit{Collection Constraint Analysis Prompt}. These prompts instruct the LLM to re-evaluate the Analyzer results sequentially: first assessing the semantic coherence of each association (e.g., if the policy mentions cloud synchronization but the result identifies the data type as ``Device information,'' the entry is marked inconsistent and discarded); then removing duplicate entries; and finally verifying compliance with the official DSS disclosure constraints. Specifically, the model checks whether the declared \textit{data type} and its \textit{policy reference} are consistent.

Ultimately, the \textit{LLM Post-Processing Module} classifies the refined results into two categories. 
The first, \textit{omitted declarations}, represents confirmed inconsistencies where the Privacy Policy declares data collection, sharing, or other practices not reflected in the corresponding DSS entry. 
The second, \textit{excluded declarations}, captures initially detected inconsistencies that, after post-processing validation against DSS disclosure rules, are identified as false positives or non-violations. These cases arise when a finding appears as an omission at first glance, but either (i) its policy reference pertains to a different data-handling purpose that does not require DSS disclosure (e.g., the result marks “Approximate Location” as omitted, but the related text concerns “User Payment Information”), or (ii) the data type and policy reference fall outside Google Play’s defined DSS disclosure constraints and therefore do not require declaration.

The output structure in Listing~\ref{lst:json-output2} is similar to that of Listing~\ref{lst:json-output}, with the addition of an excluded declarations section. This section provides, for each exclusion, the underlying reason, including: (i) \texttt{reason\_of\_removal}, a keyword summarizing the reason, and (ii) \texttt{justification}, which briefly explains the reason for removal.
\begin{lstlisting}[ caption={LLM Post-Processing JSON output structure.}, label={lst:json-output2},  style=privacycode, escapeinside={(*@}{@*)}, numbers=none]
{
    "omitted_declarations": [
        {
            "data_type": "Data type",
            "policy_reference": "Exact excerpt from the privacy policy",
            "lang": "PP language"
        }
    ],
    "excluded_declarations": [
        {
            "data_type": "Data type",
            "policy_reference": "Exact excerpt from the privacy policy",
            "reason_of_removal": "Keyword explaining why Analyzer finding is not a true DSS omission",
            "justification": "Brief reasoning why not a true omission, referencing DSS categories",
            "lang": "PP language"
        }
    ]    
}
\end{lstlisting}

\section{Implementation}
\label{sec:implementation}

In this section, we explain how we implemented \toolName and the reason behind the technical choices. 

\paragraph{Scraper Module.}The scraper module is composed of the \textit{Data Safety Extractor}, a Node.js component that relies on the open-source \texttt{google-play-scraper} tool available on GitHub.\footnote{https://github.com/walkerliu01/google-play-scraper}.Given a package name, it extracts all the details present in the DSS and stores them in a structured JSON file.
The \textit{Privacy Policy Extractor} is a Python-based module that takes as input the Privacy Policy URL identified by the \textit{Data Safety Extractor} and retrieves the corresponding policy content. The module launches a headless browser, sets a custom \texttt{User-Agent} to emulate realistic user behavior, and navigates to the specified URL. After a configurable waiting period of 3 seconds—introduced to allow dynamic JavaScript content to fully load—the rendered HTML content is collected.

A key challenge in this process is the presence of cookie consent banners, which may obscure or block access to the Privacy Policy content. To address this issue, we adopted a twofold strategy. First, we defined a set of heuristic rules based on commonly observed consent elements, including keywords and HTML selectors (e.g., \texttt{button:has-text('ACCEPT')}, \texttt{button:has-text('AGREE')},\texttt{button:has-text('REJECT')}). 
Second, we evaluated the effectiveness of browser extensions designed to automatically handle cookie banners. In particular, we considered \textit{I don't care about cookies}\footnote{https://chromewebstore.google.com/detail/i-dont-care-about-cookies/fihnjjcciajhdojfnbdddfaoknhalnja?pli=1}, \textit{uBlock}\footnote{https://chromewebstore.google.com/detail/ublock/epcnnfbjfcgphgdmggkamkmgojdagdnn?hl=en} and \textit{Consent-O-Matic}\footnote{https://chromewebstore.google.com/detail/consent-o-matic/mdjildafknihdffpkfmmpnpoiajfjnjd}. The evaluation was conducted on a randomly selected sample of 20 applications, with the goal of assessing their ability to consistently expose the PP content without manual intervention. The results indicated that \textit{I don't care about cookies} achieved higher effectiveness in bypassing or removing cookie banners across diverse websites.
To further validate this choice, we extended the evaluation of \textit{I don't care about cookies} to the full dataset of 330 applications. The results show that, in the vast majority of cases, the extension successfully handled cookie consent banners without requiring manual interaction. Specifically, testing the extension on 330 apps, only 10 of them (approximately 3\%) still presented residual cookie banners that could potentially interfere with the extraction process. This confirms the robustness of the adopted solution in real-world conditions, while also highlighting a limited set of edge cases where additional handling strategies may still be required.
Based on this empirical assessment, we integrated this extension into the Chromium-based scraping pipeline.

However, this component is mainly meant to support our experimental campaign and automatically collect the privacy policy for many apps. The market provider who needs to check apps before accepting them, or development teams willing to check their apps, typically have direct access to privacy policies with no complication caused by cookie banners.

% However, since web page structures vary, the removal process is not always successful. In those cases, the PP Web Page may be partially obscured by the cookie banner, making the LLM's PP analysis much more difficult and sometimes ineffective. \lv{questa frase co condanna un pò dobbiamo girarla al positivo dicendo per minimizzare il rischio di cookie banner abbiamo usato le strategie sopra; poi questa cosa che potrebbero parzialmente essere oscusate forse lo farei nella discussione, magari se avessimo dei dati su questi oscuramenti nei nostri test sarebbe ancora meglio} 

% Once the page finishes loading and the cleaning attempt is completed, the fully rendered content is saved as a PDF file. We opted for the PDF format, rather than extracting the text through libraries such as \textit{BeautifulSoup}, because in our analysis we employ \textbf{Gemini 2.5-pro} and the \textbf{Gemini API Document Understanding}~\cite{}. As explained in the official documentation, Gemini uses \textit{native vision} for PDF files to analyze text, images, tables, and layout (up to 1000 pages), whereas “non-PDF documents” such as plain text files are treated as regular text, thus losing the visual context and formatting.

\paragraph{LLM Analysis Modules.} All LLM Modules, namely LLM Pre-Processing, LLM Analyzer, and LLM Post-Processing, are implemented as Python scripts. For each module, we leverage a \textit{Zero-shot Prompting} strategy, as defined by Zhou et al.~\cite{lo}.
Regarding the choice of the LLM, we compare the following state-of-the-art models: i)\textbf{ Gemini 2.5-pro} ii)\textbf{ deepseek-v3.1:671b-cloud} iii) \textbf{qwen3:8b}.
In addition, we conducted an extensive prompt engineering study to identify the most effective prompting strategy for our pipeline. We report the results of our analysis in Section~\ref{sec:ExpRes}. Listings~\ref{lst:llm-preproc}-\ref{lst:llm-main}-\ref{lst:llm-postproc} in Appendix~\ref{Prompt} report the structure of the resulting prompts for pre-processing, analysis, and post-processing, in accordance with the strategies discussed in Section~\ref{sec:methodology}.

% \paragraph{Merge Results Module.} This module consists of two Python scripts. The first one, the \textit{Refinement Results Script}, is responsible for removing the first and last lines from the LLM Analyzer output. This step is required because the JSON results generated by the LLM are enclosed between the markers \texttt{'''json} at the beginning and \texttt{'''} at the end. If these lines are not removed, the file cannot be properly parsed, resulting in an “invalid JSON” error during the upload process.
% The second script, the \textit{Merge Results Script}, merges the three collection results (i.e., Omitted UserData Declarations, Omitted TechnicalData Declarations, and Omitted Location Declarations) and the three share results (i.e., Omitted UserData Declarations, Omitted TechnicalData Declarations, and Omitted Location Declarations) obtained by the \textit{LLM Analyzer} into a single JSON file for collection results and a single JSON file for share results (i.e., Refined Omitted Collection Declarations and Refined Omitted Share Declarations).

\section{Experimental Campaign and Analysis Results}
\label{sec:ExpRes}
To answer the research questions introduced in Section~\ref{sec:intro} (RQ1–RQ3), we adopt a two-phase experimental methodology. In both phases, our analysis operates exclusively on application package names and publicly available metadata collected from the Google Play Store. Package names are automatically identified and extracted from the official Google Play Store pages using a custom Selenium-based script; no application binaries are downloaded or analyzed.

In the first phase, we conduct a pilot study on a benchmark of 10 top-ranked Google Play Store applications. This phase is used to systematically tune the experimental parameters of \toolName, including prompt configurations and the choice of the underlying LLM. The objective of this pilot is to select a stable and effective configuration, which serves as the baseline for all subsequent experiments.

In the second phase, we apply \toolName with the finalized configuration to a large-scale empirical analysis (RQ1). We collect the package names of the top-ranked application in Q3 2025 from each of the 33 Google Play Store categories, resulting in a dataset of 330 applications.
This phase enables a quantitative assessment of omission prevalence across both individual data types (RQ2) and aggregated data categories (RQ3). 
The experimental campaign was carried out on a 12th Gen Intel(R) Core(TM) i9-12900KS (3.40 GHz) server with 128 GB RAM, and a 24 GB NVIDIA GeForce RTX 4090. 
The results of this analysis are presented in the following sections.

\subsection{Experimental Setting}

\paragraph{Metrics.} 
For our analysis, we define True Positives (TPs) as the omitted declarations detected by \toolName that are present in the PP but not in the DSS.
We classify as False Positives (FPs) omitted declarations identified by \toolName that are inconsistent in terms of:  policy reference results, and/or non-compliance with the Google Data Safety constraints. 
True Negatives (TNs) refer to omitted declarations correctly classified as excluded from \toolName.
Finally, we classify as False Negatives (FNs), all the omitted declarations that \toolName fails to identify during the analysis and the excluded declarations that are misclassified by the LLM Post-Processing Analyzer.
Based on this classification, we compute four standard evaluation metrics:
\begin{itemize}
    \item Precision: TP/(TP+FP).
    \item Accuracy: (TP+TN)/(TP+TN+FP+FN).
    \item Recall: TP/(TP+FN).
    \item F1-Score: 2*[(Precision*Recall)/(Precision+Recall)].
\end{itemize}

% \mc{presenta e spiega le metriche che misuri (es. TP, FP, F1, ...) dandone la definzione (es con la formula usata) }

\paragraph{Datasets.} \label{dataset}
The experimental methodology is supported by three distinct datasets.
%\textcolor{red}{The first, $set_{small}$, includes 10 apps randomly selected from the top 100 chart of the Google Play Store in Q3 2025 and is used to identify the best prompt strategy and the most suitable LLM model.}

The first, $set_{small}$, consists of 10 apps randomly drawn from the top chart of the Google Play Store in Q3~2025. It is used exclusively to identify the best prompt configuration and the choice of LLM model, and is entirely held out from the subsequent empirical evaluation. To reliably estimate false negatives during this phase, the privacy policies of all 10 apps
were manually reviewed in full.

% \textcolor{blue}{ Non fa parte della validazione sperimentale   abbiamo approccio  ---> congeliamo approccio ---> verifichiamo sperimentalmente le ipotesi definite nell'approcio}
The second, $set_{large}$, contains a collection of 330 package names associated with apps published on the Google Play Store with a minimum of 50,000 installations, spanning all 33 store categories (i.e., 10 apps per category). $set_{large}$ is used for the large-scale empirical analysis.

Finally, $set_{medium}$ is a stratified 10\% random sample of $set_{large}$, containing one app per Google Play Store category for a total of 33~apps. Stratified sampling ensures that every category is represented. $set_{medium}$ is used exclusively for manual ground-truth
annotation and performance estimation.

%extracted 10 apps for each Google Play Store category of top-zero dollar application for a total of 330 apps, with a minimum of 50,000+ installations, and $set_{medium}$, which is a 10\% random sample of $set_{large}$ that maintains all the app categories, so we have 33 apps, one for each categories.
%While previous studies have analyzed millions of apps with superficial checks, we opted for a stratified dataset of 330 apps to allow for a complete manual Deep Validation (Ground Truth), necessary to demonstrate the effectiveness of the methodology.
% \mc{tra i case study set, manca il gruppo di 10 app usate per il tuning: $set_{small}$?}

% \mc{presenta i due set di casi di studio usati (10 app + 330 app) spiegando come le hai prese e perche' secondo noi sono rappresentative. Dai un nome ai due data set (small-set e large-set, o set$_1$ e set$_2$ ) e poi referenzia questo nome.}

\paragraph{Experimental procedure.}
We applied \toolName to the selected datasets following the two-phase experimental methodology described above and collected all the reported inconsistencies.
During the pilot phase, \toolName was applied to $set_{small}$ to support the selection of the most effective prompt configuration and LLM model. In the large-scale phase, the finalized configuration was applied to $set_{large}$ to perform the empirical analysis.

To validate our findings, we conducted a manual analysis on $set_{medium}$, a representative subset of $set_{large}$, to ensure strict coherence between the LLM-generated outputs and the corresponding source material. Specifically, we cross-verified that the privacy policy reference citations reported in the results accurately matched the text of the previously scraped PP. 
Furthermore, we manually confirmed that each Data Type labeled as either omitted or excluded was consistent with the content of the referenced PP excerpts.
This verification process enabled us to reliably classify each outcome as a True Positive (TP), False Positive (FP), True Negative (TN), or False Negative (FN), and to subsequently compute Precision, Accuracy, Recall, and F1-score.

Finally, since the non-deterministic behavior of LLMs is a well-known issue in the literature~\cite{tosem25}, we assessed the robustness, correctness, and reproducibility of our approach by executing the full analysis pipeline three times on $set_{medium}$ and comparing the obtained results. 
All models were evaluated using a temperature setting of 0 to reduce randomness in the generation process and limit output variability across runs. While this configuration does not ensure complete determinism, it helps reduce randomness and enables us to focus the analysis on the inherent differences between models, rather than variations caused by sampling.

\paragraph{Ground truth annotation.}
\label{sec:groundtruth}

Ground-truth labels for $set_{medium}$ were produced by a single annotator following a structured annotation protocol. For each of the 33 apps, the annotator independently examined the full PP text and the corresponding DSS as scraped from Google Play. This review was conducted without prior access to \toolName's outputs to avoid automation bias.

The annotation process consisted of the following steps: 
(i) manual reading of the entire PP text, with explicit identification and marking of statements describing data collection and data sharing practices; 
(ii) mapping of each identified statement to one of the 38 data types defined in the Google Play taxonomy; 
(iii) comparison of these manually derived labels against the corresponding declarations in the DSS; and 
(iv) labeling of each PP--DSS pair as either a complete match or an omission (i.e., a discrepancy between declared and observed practices).

For each data type, the annotator determined whether the app collects or shares that information according to the PP, and then assessed consistency with the DSS. Each case was classified as a True Positive (TP) if \toolName correctly flagged an omission, a False Negative (FN) if it missed it, a True Negative (TN) if it correctly identified a compliant case under Google's disclosure constraints, or a False Positive (FP) if it incorrectly flagged a non-issue.

In total, the annotation covered 108 sharing omissions and 247 collection omissions, for a total of 355 cases across the 33 apps in $set_{medium}$. As part of the same process, the annotator also verified that the PP excerpts extracted by the Pre-Processing Module accurately reproduced the original policy text.
Since only one annotator performed the labeling, a degree of subjectivity cannot be fully excluded. However, the task is constrained by Google's well-defined data-type taxonomy and DSS disclosure rules, which substantially reduce interpretative ambiguity.

% \textcolor{red}{!!!!!!!!!!!!!!}

 While in principle its high level of automation would allow PolicyGapper to process a much larger set of apps, the experiment requires detailed and accurate manual analysis of the policy text %by independent reviewers 
to identify all the policy inconsistencies. This highly time consuming gold standard preparation step represents the main constraint to the size of the dataset.

% \paragraph{Experimental procedure.}
% We applied \toolName to the case studies and collected all the reported inconsistencies.
% To validate our findings, we manually analyzed the results to ensure strict coherence between the LLM-generated output and the source material. We cross-verified that the privacy policy reference citations in the results matched the text of the previously scraped PP. Furthermore, we manually confirmed that any Data Type declared as either "omitted" or "excluded" was consistent with the content of the referenced PP text cited in the result. This rigorous verification process allowed us to accurately define each result as TP, FP, TN, or FN, and, as a consequence, to evaluate Precision, Accuracy, Recall, and F1-Score.
% In addition, since in the literature the problem related to the non-deterministic results of the LLMs\cite{tosem25} is well known, to assess the robustness, correctness, and reproducibility of our approach, we executed the full pipeline three times, on $set_{medium}$.

% \mc{spiega la procedura usata per raccogliere i valori delle metriche che hai identificato} 

%To define the final settings of the empirical validation, we run a small-scale experiment to compare alternative configurations, they are how to split the prompt and which model to use.

\subsection{Pilot Study}

The pilot study focused on defining the best settings for \toolName in terms of the LLM model used and the configuration of the prompts, with respect to their number and contextual scope.

\paragraph{Prompt engineering:}
% \mc{in problema affrontato non e' chairo: spiega che cosa vuol dire splittare il prompt e perche' e' necessario}
A critical challenge in LLM-based privacy analysis is managing the trade-off between instruction density and model attention. Liu et al.~\cite{liu-etal-2024-lost} show that model performance degrades when relevant information is placed in the middle of longer contexts. Conversely, querying the model too frequently for individual data types can increase the risk of hallucinations.

To ensure experimental control and avoid cross-query interference, we enforce stateless interactions with the LLM. Each prompt is executed in isolation, without retaining any prior conversational context or intermediate outputs. This guarantees that each analysis step remains independent and prevents unintended information carryover across queries.

Building on this controlled setup, we evaluated four prompt decomposition strategies using Gemini 2.5 Pro, exploring different levels of granularity, ranging from a unified prompt to a fine-grained, component-based approach:
\begin{itemize}
\item (i) a single comprehensive prompt covering all Data Types; 
\item (ii) 3 prompts, each focused on a group of related Data Categories (Device Data, User Data, and User-Generated Data); 
\item (iii) 14 prompts, one per Data Category; and 
\item (iv) 38 prompts, one per Data Type. 
\end{itemize}

 % \mc{spiega il contenuto della tabella}
Among the considered setups, as shown in Table \ref{tab:promptSelection}, the three-prompt configuration achieved the best overall performance on $set_{small}$.
The single-prompt setup reduced coverage, leading to approximately a 50\% increase in false negatives compared to the multi-prompt variants. Conversely, the 14-prompt and 38-prompt configurations caused excessive overfitting and keyword-based matching, notably inflating false positives—56 and 72 FPs, respectively, versus only 15 FPs with the three-prompt setup. This behavior indicates that while finer granularity promotes specificity, it limits the model’s semantic generalization capabilities. Based on these results, the three-prompt configuration was selected for subsequent analyses, as it provides an optimal balance between precision and recall in this evaluation setting.
\begin{table}[t]
\centering
\scriptsize
\definecolor{geminigreen}{HTML}{B2FFB9}

\begin{tabular}{lcccc}
\toprule
\textbf{Metrics / \# of Prompt} & \textbf{1} & \cellcolor{geminigreen}\textbf{3} & \textbf{14} & \textbf{38} \\
\midrule
Precision & 0.68 & \cellcolor{geminigreen}\textbf{0.84} & 0.59 & 0.52 \\
Accuracy  & 0.53 & \cellcolor{geminigreen}\textbf{0.71} & 0.61 & 0.58 \\
Recall    & 0.58 & \cellcolor{geminigreen}\textbf{0.71} & 0.71 & 0.71 \\
F1-Score  & 0.62 & \cellcolor{geminigreen}\textbf{0.77} & 0.65 & 0.60 \\
\bottomrule
\end{tabular}

\caption{Performance of the four prompt configurations (1, 3, 14, and 38 prompts) on the $set_{small}$ for detecting omitted declarations. The 3-prompt configuration achieves the best overall performance and is highlighted in green.}
\label{tab:promptSelection}
\end{table}

\paragraph{Model Selection:}
After selecting the three-prompt configuration as the most effective setup, we conducted a second experiment to identify the most appropriate model to integrate into our pipeline. We selected two freely accessible models and one paid model, namely \texttt{Qwen3:8B}, \texttt{DeepSeek-V3.1:671B-Cloud}, and \texttt{Gemini~2.5~Pro}. 

The selection criteria were defined to cover three main paradigms relevant to research and automated analysis:
\begin{itemize}
    \item Local open-weights (\texttt{Qwen3:8B}): represents a fully privacy-preserving approach, running entirely on local hardware to assess feasibility without external data transmission.
    
    \item Cloud open-weights (\texttt{DeepSeek-V3.1}): represents high-performance open models accessed via API, combining the transparency of open weights with the computational power of cloud infrastructure.
    
    \item Cloud proprietary (\texttt{Gemini 2.5 Pro}): represents the commercial 
    state of the art, offering closed-source, multimodal capabilities managed entirely by the provider. Each API request is executed in a stateless manner, ensuring that no contextual information is retained across different applications and preventing cross-analysis contamination.
\end{itemize}

Beyond performance alone, the choice of the final model is guided by a combination of practical and technical factors that are critical for our task. In particular, privacy policy analysis requires the ability to process long, structured documents, preserve layout information (e.g., tables, sections), and maintain consistency across independent analyses. 

Both Gemini 2.5 Pro and DeepSeek-V3.1:671B-Cloud run as cloud-based models, while Qwen3:8B runs on our local hardware. We attempted to run models with more than 8B parameters locally, but their computational cost did not scale on our hardware, making the analysis infeasible in terms of execution time. 

The evaluation was performed on $set_{small}$, for which the PPs were manually reviewed to reliably compute FNs. Both Qwen3:8B and DeepSeek-V3.1:671B-Cloud do not support direct file uploads, whereas Gemini 2.5 Pro supports this functionality. As stated in the official Gemini API documentation~\cite{Gemini}, Gemini 2.5 Pro supports direct file uploads and implements \emph{native vision} for PDF processing, enabling comprehensive analysis of text, images, tables, and layouts in documents up to 1{,}000 pages. This capability is particularly relevant in our context, where privacy policies often rely on structural cues (e.g., headings, formatting, tabular disclosures) that are partially lost when converted to plain text.

In contrast, non-PDF formats such as plain\texttt{.txt} files are treated as unstructured text, thereby losing their visual and formatting context. To assess this difference, we compared Gemini 2.5 Pro's performance on equivalent policy content provided in both PDF and TXT formats. To ensure a fair comparison across all three models, and to avoid relying on direct PDF uploads, we used the Python library PyPDF2\footnote{\url{https://pypi.org/project/PyPDF2/}} to extract text from PDF files and embed it directly into the prompt for analysis.

Experimental results are shown in Table~\ref{tab:LLMSelection}. Gemini 2.5 Pro consistently achieves the best performance across all key metrics, reaching the highest Precision (0.84), Accuracy (0.71), Recall (0.71), and F1-score (0.77). 

However, the superiority of Gemini 2.5 Pro is not limited to raw performance. Compared to the other models, it exhibits:
(i) greater robustness, with no observed pipeline failures or malformed outputs,
(ii) better adherence to structured output constraints (e.g., JSON format),
and (iii) more stable behavior across runs, which is essential for reproducible analysis.

While Qwen3:8B exhibits the fastest inference time (256.5\,s), its reliability is severely compromised, failing in approximately 40\% of cases due to pipeline crashes, constraint violations, and out-of-context outputs. Gemini 2.5 Pro without file uploads achieves a competitive Recall (0.35) compared to the other models that do not support file uploads, but exhibits slower execution (383.6\,s) and incurs a minor cost (\$ 0.50 per run).

When enabling file uploads (PDF version of the PP, DSS JSON, and additional artifacts), Gemini 2.5 Pro further improves its effectiveness, consistently outperforming all other configurations on the $set_{small}$ dataset. In this setting, it achieves a 180\% improvement in Precision, 144.83\% in Accuracy, 102.86\% in Recall, and 140.62\% in F1-score with respect to the version without file uploads. 

This improvement can be attributed to the model’s ability to directly process structured documents without flattening them into raw text, thus preserving semantic and layout information. Furthermore, the use of file uploads reduces response latency compared to embedding raw text directly in the prompt. As noted in the Gemini API documentation~\cite{geminiLatency}, while every request incurs a fixed latency overhead, total latency generally increases with query length. Consequently, passing document content directly within the prompt results in a substantially larger payload than referencing uploaded files, thereby leading to slower processing times.

Overall, these results suggest that Gemini 2.5 Pro represents a particularly suitable choice for our pipeline, as it combines strong analytical performance with practical advantages in terms of robustness, input handling, and scalability. While this does not constitute a definitive conclusion on model superiority, it provides promising evidence that multimodal, document-aware LLMs can significantly enhance privacy policy analysis.

% Concerning the results of uploading the .pdf PP Web Page, the .json DSS, and all the other results analysis files, Gemini 2.5 pro, leveraging \emph{native vision} for PDF processing, outperforms all the other analysis results on the $set_{small}$ dataset.
% Gemini 2.5 pro, with the upload of the files, has a 180\% improvement for the Precision metrics, 144.83\% for the Accuracy, 102,86\% for the Recall, and 140.62\% for the F1-Score with respect to the version without the upload of the files. 
% Furthermore, utilizing the file upload feature with Gemini 2.5 Pro reduces response latency compared to embedding raw text. As noted in the Gemini API documentation \cite{geminiLatency}, while every request incurs a fixed latency overhead, total latency generally increases with query length. Consequently, passing document content directly within the prompt creates a significantly larger payload than referencing uploaded files, resulting in slower processing times.
% \mc{che senso ha separare le tabelle 2 e 3, siccome sono le stesse analisi direi di unierle} 

\begin{table}[htb]
\centering
\scriptsize
\definecolor{geminigreen}{HTML}{B2FFB9}
\begin{tabular}{lcccc}
\toprule
\textbf{Metrics/Models} & \textbf{Qwen3} & \textbf{Deepseek} & \textbf{Gemini (txt)} & \textbf{Gemini (PDF)} \\
\midrule
\textbf{Precision}     & 0.29  & 0.31  & 0.30  & \cellcolor{geminigreen}\textbf{0.84} \\
\textbf{Accuracy}      & 0.25  & 0.38  & 0.29  & \cellcolor{geminigreen}\textbf{0.71} \\
\textbf{Recall}        & 0.23  & 0.41  & 0.35  & \cellcolor{geminigreen}\textbf{0.71} \\
\textbf{F1-Score}      & 0.26  & 0.35  & 0.32  & \cellcolor{geminigreen}\textbf{0.77} \\
\textbf{Avg. Time (s)} &\cellcolor{geminigreen}\textbf{ 256.5 }& 303.5 & 383.6 & 357.8\\
\textbf{Cost (\$)}     &\cellcolor{geminigreen}\textbf{ 0}     & \cellcolor{geminigreen}\textbf{0}     & 0.50  & 0.58 \\
\bottomrule
\end{tabular}
\caption{Comparison of LLM configurations on the $set_{small}$. Gemini (PDF) achieves the best trade-off between effectiveness (Precision, Accuracy, Recall, F1-score) and efficiency (time, cost).}
\label{tab:LLMSelection}

\end{table}

%This $set_{small}$ benchmark is only used to select the best prompt configuration and LLM model. All results reported for RQ1, RQ2, and RQ3 are computed on the $set_{large}$ and $set_{medium}$.
% \mc{sposta la descrizione del data set all'inizio (nuova sottosezione) e dagli un nome da referenziare qui}

\subsection{RQ1: Analysis of Detection Accuracy}

We executed \toolName on $set_{large}$ using the configuration selected during the pilot study.
Across these 330 apps, we identified a total of 2{,}689 omitted declarations, corresponding to an average of approximately 8 omissions per app. These omissions consist of 2{,}040 omitted declarations related to data collection practices and 649 omitted declarations related to data sharing practices. These represent data types marked as collected or shared in the PP but not declared in the DSS.

% Once the best configuration was obtained, we started to work in order to address RQ1.
% We analyzed the $set_{large}$ defined in~\ref{dataset},
% across these 330 apps, we identified a total of 2,040 Omitted Collection Declarations and 649 Omitted Share Declarations, amounting to 2,689 total omitted declarations, an average of 8 omissions per app. These represent data types marked as Collected or Shared in the PP but not declared in the DSS.
% \begin{table}[htb]
% \centering
% \arrayrulecolor{black} % Assicura bordi neri

% \begin{tabular}{|l|c|c|c|}
% \cline{2-4}
% \multicolumn{1}{c|}{} & TP & FP & TN \\ \hline
% Omitted Collection Declarations & 2040 & 495 & 274 \\ \hline
% Omitted Share Declarations      & 649  & 417 & 824 \\ \hline
% \textbf{Omitted Total Declarations}      & \textbf{2689} & \textbf{912} & \textbf{1098} \\ \hline
% \end{tabular}

% \caption{330 Apps Analysis Result}
% \label{tab:330Res}
% \end{table}

To accurately estimate the number of FNs, we manually analyzed the results related to the apps included in $set_{medium}$, reviewing all undeclared data types in the corresponding DSS. In addition, we executed the full pipeline two additional times on the apps in $set_{medium}$ and repeated the manual verification. These results allow us to derive the mean and standard deviation values of TPs, FPs, TNs, and FNs for all three runs (see Table~\ref{tab:statsRes}) and to compute the metrics reported in Table~\ref{tab:statsRes2}. Averaging across the three experimental runs, we recorded  $272.7 \pm 20.5$ TPs, $90.7 \pm 12.0$  FPs, $102.7 \pm 8.7$ TNs, and $95.3 \pm 2.6$  FNs, yielding an Precision of $0.75 \pm 0.04$, Accuracy of $0.69 \pm 0.07$, Recall of  $0.77 \pm 0.07$, and an F1-score of $0.76 \pm 0.04$ . 

Overall, the results indicate a reasonably stable performance across runs, with standard deviation remaining below 10\% of the mean for most metrics. In particular, the low variability observed for FNs suggests that the pipeline is consistent in identifying undeclared data types, which is critical for this task.

The achieved Recall ($0.77 \pm 0.07$) is higher than Precision ($0.69 \pm 0.07$), indicating that the approach is more effective at detecting omissions than at avoiding false positives. This behavior is desirable in our setting, where missing a potential discrepancy (i.e., a false negative) is generally more critical than flagging an additional false positive.

The slightly higher variability observed in Precision is consistent with the fluctuations in false positives across runs, likely due to the non-deterministic nature of the pipeline. Nevertheless, the overall F1-score ($0.76 \pm 0.04$) confirms a balanced trade-off between Precision and Recall, supporting the effectiveness of the proposed approach.

\begin{table}[t]
\centering

\begin{tabular}{lc}
\toprule
\textbf{Metric} & \textbf{Value} \\
\midrule
TP & $272.7 \pm 20.5$ \\
FP & $90.7 \pm 12.0$  \\
TN & $102.7 \pm 8.7$  \\
FN & $95.3 \pm 2.6$   \\
\bottomrule
\end{tabular}
\caption{Results across three runs (mean $\pm$ standard deviation) on 
$set_{medium}$.}
\label{tab:statsRes}
\end{table}

\begin{table}[t]
\centering
\begin{tabular}{lc}
\toprule
\textbf{Metric} & \textbf{Value} \\
\midrule
Precision  & $0.75 \pm 0.04$ \\
Accuracy & $0.69 \pm 0.07$ \\
Recall    & $0.77 \pm 0.07$ \\
F1-score  & $0.76 \pm 0.04$ \\
\bottomrule
\end{tabular}
\caption{Performance metrics across three runs (mean $\pm$ standard deviation) on $set_{medium}$.}
\label{tab:statsRes2}

\end{table}

\begin{tcolorbox}[colback=gray!5,colframe=blue!40, boxrule=0.3mm, arc=0mm, left=1mm, right=1mm, top=1mm, bottom=1mm]
\textbf{RQ1: Can LLMs effectively identify omitted declarations between the PP and the DSS?}
As shown by the analysis results obtained from 330 applications collected from the Google Play Store, our methodology successfully identifies 2{,}689 omitted declarations between the PP and the DSS. 
In the manually verified control dataset ($set_{medium}$), our approach achieved an average Precision of 0.75, Accuracy of 0.69, Recall of 0.77, and an F1-score of 0.76 across three runs. These results provide encouraging evidence of the effectiveness of the proposed approach.
%In the manually verified control dataset ($set_{medium}$), our approach achieved an average Precision of 0.75,  Accuracy of 0.69,
%Recall of 0.77, and F1-score of 0.76 and F1-score of 0.76 across three runs, encouraging evidence of the effectiveness of the proposed approach, which could benefit from further empirical investigation. 
\end{tcolorbox}

\subsection{RQ2: Analysis of Omitted Declarations}
To address RQ2, we explore in more detail the results obtained from the previous analysis, focusing on which declarations are most frequently omitted. The results of this analysis are shown in Table~\ref{tab:top10DT}. The most frequently omitted Data Type during the analysis is {\em Approximate Location}. This can be explained by the fact that, in most cases, when the DSS and the PP declare Data Types related to device and other identifiers—such as IMEI numbers, MAC addresses, Firebase installation IDs, advertising identifiers, or IP addresses—developers often neglect to declare Approximate Location. This omission likely occurs because the IP address itself can implicitly reveal the user’s approximate location.

The second most omitted Data Type is {\em Web Browsing History}, with a total of 153 omissions. This is particularly relevant, as web browsing data can be highly sensitive and may reveal users’ interests, habits, or even personal circumstances. Such omissions might result from developers underestimating the privacy implications of in-app web views or embedded browsers that collect navigation patterns, often for advertising or analytics purposes.

In the third position, {\em Email Address} registers 144 omissions. This finding is significant, as email addresses are widely used for authentication, user profiling, and marketing communications. Developers may assume that collecting or processing users’ email addresses for account creation or contact forms does not require explicit disclosure, leading to underreporting in the DSS despite its inclusion in the PP.

Finally, although ranked seventh, {\em User Payment Info} also exhibits a considerable number of omissions, totaling 115 cases. This is a particularly concerning result given the sensitivity of this Data Type, as it indicates limited awareness among developers regarding the importance of properly disclosing the collection and processing of financial information.
\begin{tcolorbox}[colback=gray!5,colframe=blue!40, boxrule=0.3mm, arc=0mm, left=1mm, right=1mm, top=1mm, bottom=1mm]
\textbf{RQ2: Which types of omitted declarations appear most frequently during the analysis?}

\textit{Approximate Location} is the most frequently omitted Data Type (216 occurrences), likely because developers consider location to be inferable from identifiers such as IP addresses. It is followed by \textit{Web Browsing History} (153 occurrences) and \textit{Email Address} (144 occurrences), suggesting a limited awareness of disclosure requirements for embedded web views and user identifiers. Other notable omissions include \textit{Name}, \textit{App Interactions}, and \textit{User Payment Info}, indicating a persistent pattern of underreporting of sensitive personal and behavioral data.

\end{tcolorbox}

\subsection{RQ3: Analysis by Category}
To address RQ3, we aggregate the True Positive (TP) Data Types identified in the previous analysis into the 14 Data Categories, considering both data collection and data sharing practices.

With respect to the Data Categories most impacted by sharing omissions, all categories roughly follow a similar trend, with {\em Personal Info} (total TP: 188) being the most impacted category, followed by {\em App Activity} (total TP: 125), {\em Location} (total TP: 75), and {\em Device or Other IDs} (total TP: 66), as shown in Figure~\ref{fig:heatShare} in Appendix~\ref{image}.

Regarding the Data Categories most impacted by collection omissions, a similar pattern is observed. {\em Personal Info} (total TP: 503) is again the most impacted category, followed by {\em App Activity} (total TP: 272), {\em Location} (total TP: 235), and {\em Financial Info} (total TP: 228), as shown in Figure~\ref{fig:heatCollected} in Appendix~\ref{image}.

\begin{tcolorbox}[colback=gray!5,colframe=blue!40, boxrule=0.3mm, arc=0mm, left=1mm, right=1mm, top=1mm, bottom=1mm]
\textbf{RQ3: Which Data Categories are most impacted by omitted declarations among the Google Play Store App Categories?}

Omitted declarations are most prevalent in \textit{Personal Info}, \textit{App Activity}, and \textit{Location}, which consistently rank as the most impacted categories for both data collection and data sharing. This pattern indicates that the Data Safety Section currently fails to provide sufficient transparency into the most sensitive user data—specifically, attributes that enable fine-grained user profiling and identification.

\end{tcolorbox}

\subsection{Threats to Validity}

% \mc{non so se e' una prassi nelle conferenze di security di introdurre questo tipo di sezione}

Here we discuss the threats to the validity of our experimental campaign and how we addressed them:

Threats to {\em internal validity} impacting empirical results are primarily related to the metrics adopted to answer the research questions. To mitigate this threat, we adopted standard metrics (Precision, Recall, Accuracy, and F1-score) commonly used in classification tasks, ensuring reproducibility and comparability with similar studies. Additionally, the non-deterministic nature of LLMs represents a significant internal validity threat, as identical prompts may yield different outputs across runs. We addressed this issue by conducting three independent executions on $set_{medium}$, a subset of $set_{large}$, demonstrating consistent performance with an average Precision of 0.73, Accuracy of 0.64, Recall of 0.73, and F1-score of 0.73 across runs 2 and 3, compared to 0.80, 0.77, 0.85, and 0.82 in run 1, respectively. Furthermore, manual validation of the applications in $set_{medium}$ allowed us to accurately identify TPs, FPs, TNs, and FNs, thus reducing potential measurement bias.

Threats to {\em external validity}, impacting the generalization of our findings, are mainly due to the scope of the application datasets selected for our analysis. We mitigated these threats by considering two datasets, $set_{large}$ and its subset $set_{medium}$, both of which include applications from all Google Play Store categories. However, our findings are limited to Android applications with publicly accessible PP web pages, and with available DSS on the Google Play Store. The methodology has not been evaluated on other mobile ecosystems (e.g., Apple’s Privacy Labels), paid applications, or PPs written in languages other than English. Additionally, our analysis relies on Gemini 2.5 Pro as the primary LLM, and results may vary when adopting different models or prompting strategies.
Further experiments would be required to extend our results beyond the considered datasets, particularly to other platforms and alternative privacy compliance frameworks.

\begin{table}[t]
\centering
\scriptsize

\begin{tabular}{lc}
\toprule
\textbf{Data Types} & \textbf{\# TP Omitted Declaration} \\
\midrule
Approximate Location        & 216 \\
Web Browsing History        & 153 \\
Email Address               & 144 \\
Name                        & 136 \\
App Interactions            & 135 \\
Device and Other IDs        & 117 \\
User Payment Info           & 115 \\
Purchase History            & 115 \\
Other User Generated Content& 115 \\
Photos                      & 106 \\
\bottomrule
\end{tabular}

\caption{Top 10 data types most affected by collection and sharing omissions. For each data type, we report the number of apps for which the Privacy Policy indicates data collection while the Data Safety Section omits it on $set_{large}$.}
\label{tab:top10DT}
\end{table}

\section{Discussion and Limitations}
\label{Discussion}
% \mc{qui mi aspetterei di leggere dei casi itneressanti: app (magari blasonate) che contengono inconsistenze improtati e rilevanti, per far capire la portata dei risultati ottentuti}

Our implementation faces several practical challenges arising from both the inherent behavior of LLMs and the diverse, often ambiguous ways in which PPs are written. The first and most prominent challenge relates to the terminology and legal language used in privacy policies, which is frequently complex, exception-rich, and context-dependent. In particular, ambiguities in the description of data sharing practices often complicate the interpretation of disclosures. For example, transfers to service providers—entities processing data on behalf of developers under their direct instructions—are exempt from disclosure as “sharing” under Google Play’s rules. Such regulatory subtleties are difficult to capture automatically and can lead the LLM to misclassify compliant disclosures as omissions, thereby increasing the number of false positives.

Beyond linguistic and regulatory complexities, the non-deterministic nature of LLMs introduces additional variability in the analysis results. Even when queried with identical prompts and with the temperature set to 0, models may produce different interpretations or classifications, which in turn affects reproducibility and confidence in large-scale evaluations. To explicitly assess this effect, during the validation phase we executed the full pipeline three independent times on $set_{medium}$ and evaluated the average performance across runs. This approach allowed us to quantify variability and provide evidence of its potential, while in a real deployment scenario, additional mitigation strategies could be adopted, such as performing majority voting over several independent generations.

Additional challenges were observed in real-world deployment scenarios. For instance, apps developed by a well-known film production company, with more than 500 million installations, exhibit multiple omitted personal and financial data declarations while relying on a single PP shared across multiple products. This practice causes confusion when the analyzer searches for omissions specific to one app but encounters data practices related to others. Similarly, a popular game-based learning application with more than 50 million installations references an obsolete PP link in the Google Play Store, despite maintaining an up-to-date policy on its official website. Specifically, it points to a web page last updated in 2017, leading to inconsistent or outdated data mappings.
Moreover, PPs that conflate mobile and web platforms—for example, policies that collectively refer to both under the term “Website”—introduce numerous false negatives, since our methodology explicitly excludes data collected or shared exclusively through website usage.

We also encountered cases where PPs described only the website version of a service, leaving the app’s practices unspecified, as well as cases where the PP was inaccessible, preventing our scraper from retrieving the content.
In a limited number of cases, the PP Scraper was not able to fully close cookie banners, leaving parts of the Privacy Policy partially obscured and potentially affecting the LLM-based analysis. However, as discussed in Section \ref{sec:implementation}, this issue was largely mitigated through the integration of automated handling strategies, and thus only impacts a small set of edge cases.

These real-world inconsistencies highlight the difficulty of performing automated PP–DSS consistency validation at scale and underline the need for more standardized and machine-readable privacy documentation.

Despite these challenges, \toolName effectively detects omitted declarations in widely used applications. In fact, most apps in $set_{large}$ fall within the range of 1 million to 100 million installations, with a significant peak in the 10 million+ category (105 apps). \toolName also identifies omissions in popular streaming and social networking applications with more than 1 billion installations.
Ultimately, the detection of undeclared practices in such widely used applications highlights the practical relevance and impact of our approach, as the privacy of millions of users may be affected by inconsistencies between PPs and DSS disclosures at scale.

\section{Conclusion}
\label{sec:Conclusion}

% \mc{questa sezione contiene troppi dettagli di basso livello, mentre dovrebbe presentare la protata e la rilevanza del contributo. Ti suggerisco di rivederla secondo questo schema}
% \mc{1. riassumi il problema affrontato}
% \mc{2. riassumi ad alto livello l'approccio adottato per risolverlo}
% \mc{3. riassumi i risultati, focalizzandoti sulla portata piu' che solo sui numeri}
% \mc{4. prova a pensare a che implicazioni potrebbe avere (es. adottatto come tool per gli atuori delle app per verifica prima di pubblicare? dal maintaner del market per analisi atuomatiche?)}

This work tackles the problem of omitted declarations PPs and DSSs by proposing a multi-prompt LLM-based methodology. Building on this methodology, we introduce \toolName, the first LLM-driven tool designed to automatically identify omitted declarations between PPs and DSSs.
The results obtained on real-world Android applications demonstrate the applicability and effectiveness of the proposed approach, highlighting a still largely unresolved issue in the Google Play ecosystem. Across 330 applications, we identified 2,689 omitted declarations, with more than 95\% of the analyzed apps exhibiting at least one omission. Moreover, our evaluation confirms the feasibility of the methodology, achieving an average Precision of 0.75 across three independent runs.

From a practical perspective, \toolName can assist developers in verifying the consistency between PPs and DSSs prior to app submission, reducing the risk of rejections, removals, or enforcement actions related to privacy non-compliance. Furthermore, the proposed methodology could support automated checks within the Google Play Console, helping identify discrepancies between the two documents and notifying developers of missing Data Types in the DSS.
As future work, we plan to extend the current methodology to support Apple’s Privacy Labels by adapting the data-type taxonomy and refining prompting strategies in accordance with platform-specific guidelines. Furthermore, we aim to extend discrepancy detection to the app’s actual behaviour, e.g., by analysing whether declared security practices are effectively implemented or by assessing the consistency of declared data-processing purposes. Finally, from a methodological perspective, future empirical evaluations will incorporate a larger number of execution runs to provide a more robust statistical characterization of the model’s behaviour and further mitigate the inherent non-determinism of LLM outputs.

\section*{Ethical Considerations}
\label{sec:eth}

% This study presents several ethical considerations. Our analysis leverages publicly available Privacy Policy Pages (PPPs) and Data Safety Sections (DSS) from the Google Play Store. To minimize server burden during the scraping of 330 apps (September 2025--January 2026), we implemented strict rate-limiting (one page every 30 seconds), avoiding overload even on smaller sites.
% Transparency and accountability are prioritized: results do not accuse legal violations but highlight declarative inconsistencies under Google guidelines, aiding developers via outreach and Google disclosure.
% No end-user impact occurs, as findings inform systemic improvements without app removal or user profiling.

\paragraph{Stakeholder Identification and Impacts.}
We organize this section using a stakeholder-based analysis. We distinguish between \textit{direct stakeholders}—those who create, analyze, or manage mobile apps—and \textit{indirect stakeholders}, specifically the end-users. Direct stakeholders include our research team, app developers, and marketplace operators (e.g., Google Play). Indirect stakeholders represent the millions of users who depend on accurate privacy labels, as well as regulators (e.g., GDPR enforcement bodies) who oversee compliance.

\paragraph{Research Team.}
Our study does not involve human subjects, physical experiments, or the collection of private user data. We focus exclusively on analyzing public documents: PPs and DSSs. All analyzed materials are publicly accessible documents provided by developers on the Google Play Store. We acknowledge that automated analysis relies on LLMs, which can be non-deterministic. To ensure robustness, we mitigated this limitation through rigorous manual validation and by performing multiple test runs. Our analysis does not infer, collect, or reconstruct individual user behavior, nor does it process personal data beyond what is explicitly documented in PPs and DSSs.

\paragraph{Mobile Ecosystem Stakeholders.}
Our methodology impacts two key groups within the ecosystem:
\begin{itemize}
\item \textbf{Developers:} Developers can use \textit{PolicyGapper} to pre-validate their apps before submission, reducing the risk of rejection or penalties. We are aware that automated tools can generate false positives, which might unfairly harm a developer's reputation. To mitigate this risk, we anonymized the results in this paper and followed a responsible disclosure process, contacting 53 developers privately to verify our findings. We did not publicly identify affected applications or developers, nor did we frame the reported inconsistencies as confirmed legal violations, but rather as documentation mismatches requiring human review.
\item \textbf{Marketplace Operators:} Store operators (e.g., Google) can use this methodology to improve their existing review tools (such as Checks \cite{Checks}). This allows them to move beyond simple static analysis and perform \textit{semantic consistency checks} between the legal text and the declarative labels.
\end{itemize}

\paragraph{Privacy Community and Society.}
We align our work with core ethical principles:
\begin{itemize}
\item \textit{Beneficence:} We propose a methodology to automate the detection of missing privacy disclosures. By helping users understand data practices without needing to parse complex legal texts, we support transparency and informed consent.
\item \textit{Respect for Persons:} We respect the intellectual property of the analyzed apps and cite all prior work appropriately. Furthermore, our interactions with developers were conducted professionally, prioritizing constructive remediation over public exposure.
\item \textit{Justice:} We are releasing \textit{PolicyGapper} as open-source software to reduce disparities in resources. This allows smaller developers and researchers—who often lack large legal teams—to perform the same level of compliance verification as major corporations.
\item \textit{Respect for Law and Public Interest:} Our research promotes compliance with regulations like GDPR and app store policies. By identifying discrepancies, we contribute to a more trustworthy digital ecosystem where privacy labels accurately reflect legal obligations.
\end{itemize}

\paragraph{Publication Impacts.}
Our methodology may influence future research into LLM-based compliance checking. To reduce misuse or misinterpretation, we have clearly documented the limitations of our approach, specifically regarding the non-determinism of the results and the necessity of manual ground-truth validation.

\paragraph{Potential Harms Analysis.}
We identified and mitigated specific risks:
\begin{itemize}
\item \textit{Risk to Developers:} There is a possibility that our tool could be misused to unfairly criticize developers for minor errors or ambiguities. We mitigate this by clarifying that \textit{PolicyGapper} highlights inconsistencies in documentation, which is not the same as a legal violation.
\item \textit{Operational Impact:} To protect the websites hosting privacy policies, we configured our scraper with strict rate-limiting (one request every 357 seconds). This ensures we do not overload or crash smaller servers during data collection. We further ensured that our data collection process respected publicly accessible endpoints and did not attempt to bypass access controls, authentication mechanisms, or platform safeguards.
\item \textit{LLM Hallucinations:} Since LLMs can produce incorrect outputs (hallucinations), there is a risk of false positives. Therefore, we present \textit{PolicyGapper} as an assistive tool for human experts, rather than a standalone decision-maker, and we explicitly discourage its use as an automated enforcement or compliance authority without human oversight.
\end{itemize}

\paragraph{Decision to Conduct and Publish.}
The main goal of this project is to connect the complexity of privacy laws with the simplicity of privacy labels. We proceeded with this work because we found that helping users understand their data and assisting developers with compliance is more valuable than the minor risk of errors and before publishing, we verified our tool's metrics through manual checks. We are now sharing our code and data because we believe openness is key to making automated privacy analysis reproducible and accountable. We believe that transparently documenting both the strengths and limitations of our approach is essential to enable responsible adoption and to prevent misuse of automated privacy analysis tools.

\section*{Open Science}
\label{sec:open}
To promote transparency and reproducibility, we provide open access to \toolName,
including the datasets used in our analyses and all experimental results,
available at \toolLink.

The released artifacts are distributed under a non-commercial research license,
allowing use for academic and research purposes.

To facilitate reproducibility across heterogeneous environments, the project includes
full Docker support, enabling the execution of the entire analysis pipeline with
minimal setup effort.

\section*{Data Availability}
\label{sec:datav}
The data and code supporting the findings of this study are publicly available at
\toolLink.

The repository includes:
(i) the list of analyzed mobile applications,
(ii) all the PP and DSS analyzed,
(iii) the collected metadata and experimental results,
(iv) the prompts and outputs generated by the LLM-based analysis, and
(iv) the scripts and Docker environment required to reproduce the experiments.

All results reported in this paper can be reproduced by following the instructions
provided in the repository.

\section*{Declaration of Competing Interest}

The authors declare that they have no known competing financial interests or personal relationships that could have appeared to influence the work reported in this paper.

\section*{Funding}
\label{sec:funding}
This research did not receive any specific grant from funding agencies in the public, commercial, or not-for-profit sectors.

\section*{Declaration of generative AI and AI-assisted technologies in the manuscript preparation process}
\label{sec:ai}
During the preparation of this work, the authors used ChatGPT, Grammarly in order to: Grammar and spelling check, Paraphrase and reword. After using these tools, the authors reviewed and edited the content as needed and take full responsibility for the content of the published article.

%% -------------------------------------------------------
%% CRediT authorship statement (auto-generated from \credit above)
%% -------------------------------------------------------
\printcredits

%% -------------------------------------------------------
%% Bibliography
%% -------------------------------------------------------
\bibliographystyle{cas-model2-names}
\bibliography{bib}
%%%%%%%%%%%%%%%%%%%%%%%%%%%%%%%%%%%%%%%%%%%%%%%%%%%%%%%%%%%%%%%%%%%%%%
%% -------------------------------------------------------
%% Appendix A — Prompt templates
%% -------------------------------------------------------
\appendix

\section{LLM Analysis Prompt Templates}
\label{Prompt}

This appendix reports the prompt templates used by \toolName across the different
stages of the analysis.

\begin{lstlisting}[
  caption={Prompt used in the LLM Pre-Processing Module to extract verbatim
           data-collection/sharing statements from Privacy Policy PDFs, enforcing
           strict encoding, boundary, and output constraints.},
  label={lst:llm-preproc},
  style=privacycode,
  escapeinside={(*@}{@*)},
  numbers=none]
You are a Privacy Policy Text Extraction Specialist focused on
data-[collection/share] disclosures.
Extract and return ALL references to data [collection/share]
from the uploaded Privacy Policy Page PDF.
STEP 1: Encoding and Character Preservation
[CONSTRAINTS]
STEP 2: Data [Collection/Share] Definition
Extract ANY text Block mentioning:
[CONSTRAINTS]
STEP 3: Extraction Boundaries
[CONSTRAINTS]
STEP 4: Verbatim Extraction
[CONSTRAINTS]
STEP 5: Output Rules
Respond ONLY with extracted blocks in PDF sequence, verbatim:
[EXACT BLOCK]
If not found:
[BLANK]
NEVER add introductory text, explanations, analysis, or extra
content. Preserve original order and formatting exactly.
\end{lstlisting}

\begin{lstlisting}[
  caption={Prompt used in the LLM Analyzer Module to audit an app's PP text
           statements against its DSS, identifying undeclared data
           collection/sharing for a specified scope of review and returning a
           JSON-only discrepancy report.},
  label={lst:llm-main},
  style=privacycode,
  escapeinside={(*@}{@*)},
  numbers=none]
Act as a privacy auditor for Android apps, expert in Google
Play Data Safety.

INPUT:
SUMMARY_PRIVACY_POLICY_TXT: .txt file uploaded
DATA_SAFETY_STATEMENT (DSS) OF THE APP: .json file uploaded
SCOPE_OF_REVIEW: [DATA TO ANALYZE]

EXCLUSION CONSTRAINTS:
[DSS CONSTRAINTS]

TASK:
Analyze the SUMMARY_PRIVACY_POLICY_TXT to identify implicit or
explicit [collection/share] of SCOPE_OF_REVIEW data.
For each such data type identified, check if it is explicitly
declared as ["collected"/"share"] in the DSS.
Flag as undeclared any technical/performance data
[collected/shared] off-device but missing in DSS (outside
exclusions).

Important:
[CONSTRAINTS TO FOCUS]

RETURN ONLY valid JSON. NO extra text, comments or explanations.
Format:
[JSON RESULT FORMAT]
If none found, return exactly:
[JSON RESULT FORMAT]
\end{lstlisting}

\begin{lstlisting}[
  caption={Prompt used in the LLM Post-Processing Module to refine the raw list
           of omitted declarations: it validates consistency, removes duplicates,
           applies Google Play Data Safety exemption rules, and outputs a cleaned
           JSON of omitted and excluded declarations.},
  label={lst:llm-postproc},
  style=privacycode,
  escapeinside={(*@}{@*)},
  numbers=none]
You are a Google Play Data Safety compliance expert. Analyze the
following JSON of "omitted_declarations" from an Android app
privacy policy analysis:

STEP 1: Consistency Check
For each entry:
- Verify if "data_type" logically matches "policy_reference".
- Delete inconsistent entries.

STEP 2: Duplicates Removal
- Remove exact duplicate entries (same data_type +
  policy_reference).

STEP 3: Exemption Check
(Google Play Data Safety - [Collected/Shared] Data)
Delete entries meeting ANY exemption:
[CONSTRAINTS]

CRITICAL: Exempt ONLY if policy EXPLICITLY indicates
on-device-only OR proper E2EE. Generic "encrypted" or
"secure transmission" does NOT qualify.

STEP 4: Output Rules
- Respond ONLY with cleaned JSON:
[JSON FORMAT RESULT]
- NEVER add commentary, explanations, or extra text.
- Preserve original structure for remaining entries.
\end{lstlisting}

%% -------------------------------------------------------
%% Appendix B — Category-level heatmaps
%% -------------------------------------------------------
\section{Category-Level Distribution of Omitted Declarations}
\label{image}

This appendix reports the distribution of true positive omitted declarations across
app categories and Data Safety data categories, supporting the analysis of RQ3.

\vspace{1em}
\noindent\begin{minipage}{\textwidth}
  \centering
  \includegraphics[width=\textwidth]{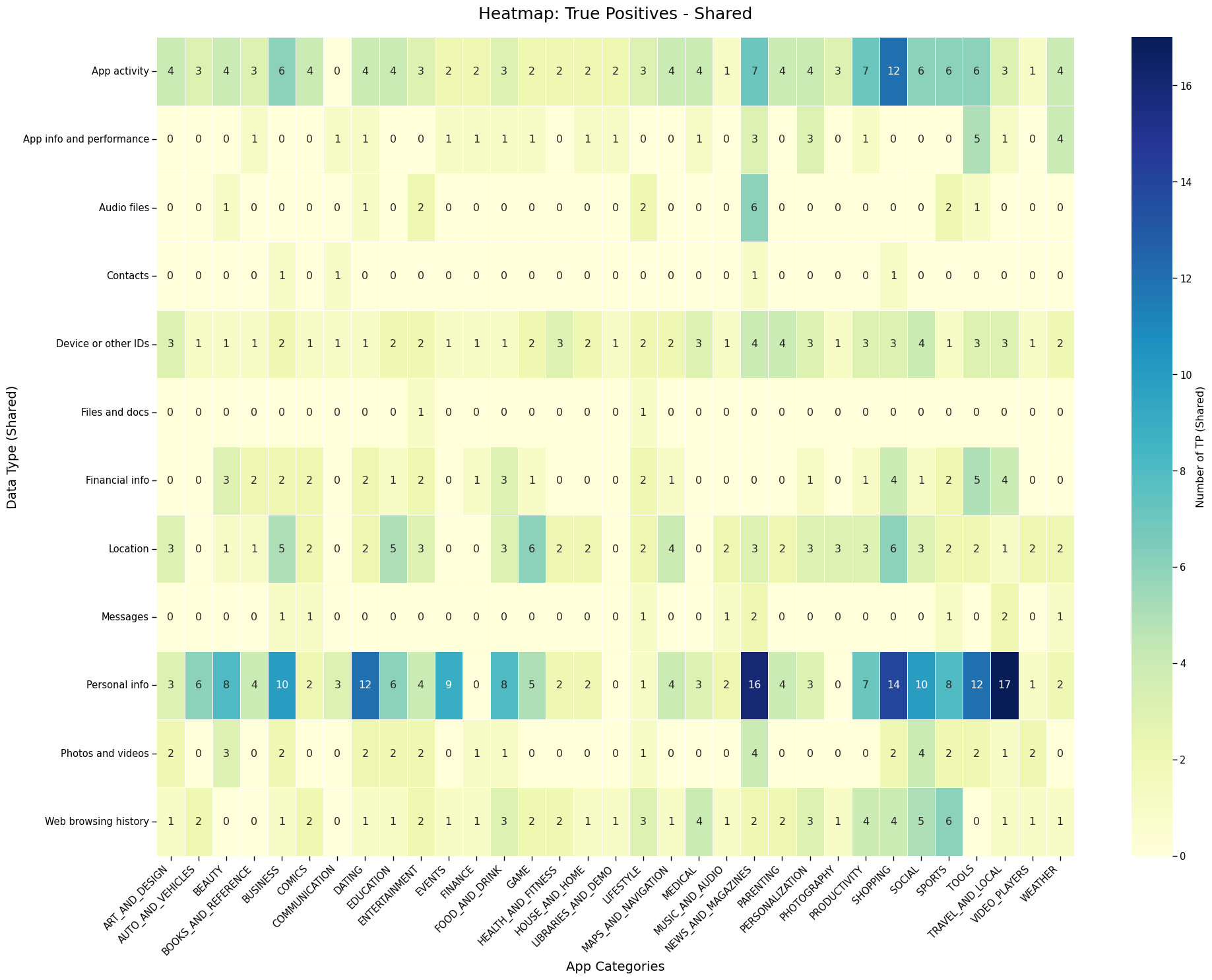}
  \captionof{figure}{Distribution of TP sharing omissions across app categories (x-axis) and
           Data Safety data categories (y-axis). Darker cells indicate a higher number
           of apps for which the PP reports data sharing while the DSS omits it.}
  \label{fig:heatShare}
\end{minipage}
\vspace{1em}

\vspace{1em}
\noindent\begin{minipage}{\textwidth}
  \centering
  \includegraphics[width=\textwidth]{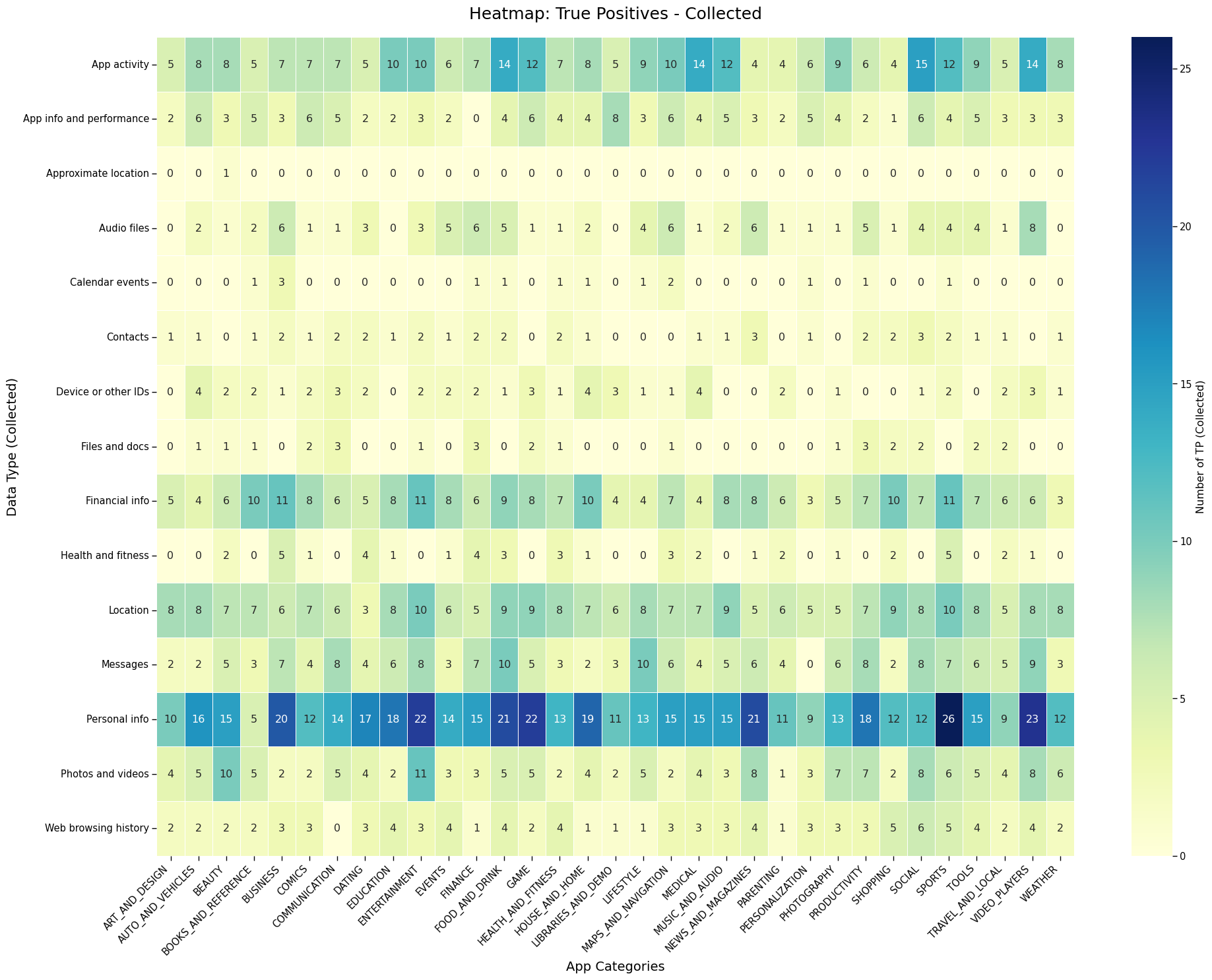}
  \captionof{figure}{Distribution of TP collection omissions across app categories (x-axis) and
           Data Safety data categories (y-axis). Darker cells indicate a higher number
           of apps for which the PP reports data collection while the DSS omits it.}
  \label{fig:heatCollected}
\end{minipage}
\vspace{1em}

\end{document}